\documentclass[sigconf]{acmart}
\AtBeginDocument{%
  \providecommand\BibTeX{{%
    \normalfont B\kern-0.5em{\scshape i\kern-0.25em b}\kern-0.8em\TeX}}}

\copyrightyear{2023}
\acmYear{2023}
\setcopyright{rightsretained}
\acmConference[KDD '23]{Proceedings of the 29th ACM SIGKDD Conference on Knowledge Discovery and Data Mining}{August 6--10, 2023}{Long Beach, CA, USA}
\acmBooktitle{Proceedings of the 29th ACM SIGKDD Conference on Knowledge Discovery and Data Mining (KDD '23), August 6--10, 2023, Long Beach, CA, USA}
\acmDOI{10.1145/3580305.3599242}
\acmISBN{979-8-4007-0103-0/23/08}

\usepackage{algorithmic}
\usepackage{graphicx}
\usepackage{textcomp}
\usepackage{pifont}
\newcommand{\cmark}{{\ding{51}}}%
\newcommand{\xmark}{{\ding{55}}}%
\usepackage{xcolor}
\definecolor{OliveGreen}{HTML}{00A64F}
\usepackage[para]{footmisc}

\usepackage[english,linesnumbered,vlined,ruled,nokwfunc]{algorithm2e}
\DontPrintSemicolon
\SetKwComment{note}{$\triangleright$ }{}
\SetFuncSty{textsc}
\SetCommentSty{textit}
\SetDataSty{texttt}
\SetKwInput{Initial}{Initial}
\SetKwInput{Parameter}{Param.}
\SetKwInput{Input}{Input}
\SetKwInput{Data}{Data}
\SetKwInput{Output}{Output}
\ResetInOut{Result} 
\let\oldnl\nl%
\newcommand{\nonl}{\renewcommand{\nl}{\let\nl\oldnl}}%
\SetKw{KwAnd}{and} %
\SetKw{KwOr}{or}
\SetKw{KwXor}{xor}
\SetKw{KwNot}{not}
\SetKw{Parallel}{parallel}
\SetKw{Return}{return}

\usepackage{graphicx}
\newcommand{\tg}{\mathcal{G}}
\newcommand{\tge}{\mathcal{E}}

\usepackage{xspace}
\newcommand{\hop}{\mathcal{H}\xspace}
\newcommand{\sh}[2]{s_{#1}^{(#2)}}
\newcommand{\hf}[3]{h_{#1,#2}^{(#3)}}
\newcommand{\hfh}[2]{h_{#1}^{(#2)}}

\newcommand{\tdegout}[2]{\delta^+(#1,#2)\xspace}
\newcommand{\tdegin}[2]{\delta^-(#1,#2)\xspace}
\newcommand{\tdegstar}[2]{\delta^\star(#1,#2)\xspace}
\newcommand{\hi}[1]{\hop(#1)\xspace}
\newcommand{\hit}[2]{\mathcal{H}_{#1}(#2)\xspace}
\newcommand{\maxdeg}{\delta_{\text{max}}\xspace}

\newcommand{\algr}{\textsc{Recurs}\xspace}
\newcommand{\algs}{\textsc{Stream}\xspace}

\usepackage{booktabs}
\usepackage{multirow}
\usepackage{multirow}
\usepackage{skull}
\usepackage{thmtools}		
\usepackage{mleftright}
\usepackage{thm-restate}
\usepackage[mathic=true]{mathtools}
\usepackage{fixmath} 

\newtheorem{theorem}{Theorem}
\newtheorem{lemma}{Lemma}
\newtheorem{definition}{Definition}
\usepackage{enumitem}
\usepackage{url}
\usepackage{subcaption}
\usepackage{cleveref}
\theoremstyle{remark}
\usepackage{multicol}

\usepackage{tikz}
\usetikzlibrary{arrows,decorations.pathmorphing}
\tikzset{
    vertex/.style={circle,thick,draw,minimum size=4mm,inner sep=0pt,font=\footnotesize},
    treevertex/.style={circle,thick,draw,fill=white,minimum size=4mm,inner sep=0.5pt,font=\scriptsize},
    edge/.style={-,thick,font=\footnotesize},
    hledge/.style={-,thick,red,font=\footnotesize},
    hlbedge/.style={-,thick,blue,font=\footnotesize},
    hlgedge/.style={-,thick,Green,font=\footnotesize}
}
\usetikzlibrary{arrows.meta}

\begin{document}

\title{A Higher-Order Temporal H-Index for Evolving Networks}

\author{Lutz Oettershagen}
\email{lutzo@kth.se}
\orcid{1234-5678-9012}
\affiliation{%
    \institution{KTH Royal Institute of Technology} %
    \city{Stockholm}
    \country{Sweden}
}

\author{Nils M.~Kriege}
\email{nils.kriege@univie.ac.at}
\orcid{0000-0003-2645-947X}
\affiliation{%
    \institution{University of Vienna} %
    \city{Vienna}
    \country{Austria}
}

\author{Petra Mutzel}
\email{petra.mutzel@cs.uni-bonn.de}
\orcid{0000-0002-2526-8762}
\affiliation{%
    \institution{University of Bonn} %
    \city{Bonn}
    \country{Germany}
}

\begin{abstract}
The H-index of a node in a static network is the maximum value $h$ such that at least $h$ of its neighbors have a degree of at least~$h$.
Recently, a generalized version, the $n$-th order H-index, was introduced, allowing to relate degree centrality, H-index, and the $k$-core of a node. We extend the $n$-th order H-index to temporal networks and define corresponding temporal centrality measures and temporal core decompositions. Our $n$-th order temporal H-index respects the reachability in temporal networks leading to node rankings, which reflect the importance of nodes in spreading processes. We derive natural decompositions of temporal networks into subgraphs with strong temporal coherence. We analyze a recursive computation scheme and develop a highly scalable streaming algorithm. Our experimental evaluation demonstrates the efficiency of our algorithms and the conceptional validity of our approach. Specifically, we show that the $n$-th order temporal H-index is a strong heuristic for identifying super-spreaders in evolving social networks and detects temporally well-connected components.
\end{abstract}



\maketitle

\section{Introduction}

Measuring the centrality of nodes is one of the fundamental operations in network analysis and has various applications~\cite{das2018study,Saxena2020}.
The H-index was originally proposed by J.\,E.~Hirsch~\cite{hirsch2005index} for measuring the productivity and impact of scientists.
It is defined as the maximum value $h$ such that at least $h$ of the scientist's publications have been cited at least $h$ times.
The corresponding centrality measure for static networks defines the H-index of a node as the maximum value $h$, such that at least $h$ of the node's neighbors have a degree of at least $h$.
Recent work has shown that the index yields a successful heuristic for finding super-spreaders in static networks~\cite{lu2016h,liu2018leveraging,gao2019weighted}.
Recently, \citet{lu2016h} formally linked the established concepts of degree centrality, H-index, and the coreness of a node by introducing the $n$-th order H-index. 
The $n$-th order H-index is defined recursively using the $\hop$ operator, which, given a multiset $S$ of integers, returns the maximum integer $i$ such that at least $i$ elements in $S$ are not smaller than $i$.
So far, the ($n$-th order) H-index has been applied primarily to static networks.

\begin{figure}
    \centering
    \begin{subfigure}{0.45\linewidth}\centering
       \includegraphics[width=0.6\linewidth]{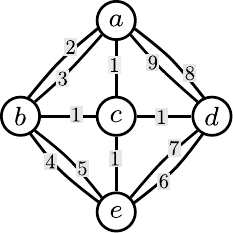}
       \caption{A small temporal network. The edges only exist on the indicated time stamps.}
       \label{fig:introexamplea}
    \end{subfigure}\hfil%
    \begin{subfigure}{0.45\linewidth}\centering
        \includegraphics[width=0.6\linewidth]{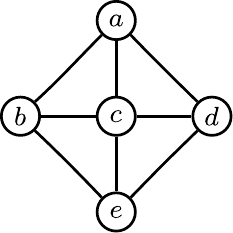}
        \caption{The underlying aggregated, static graph without temporal restrictions.}
        \label{fig:introexampleb}
    \end{subfigure}
    \begin{subfigure}{1\linewidth}\centering
        \vspace{4mm}
        \includegraphics[width=0.9\linewidth]{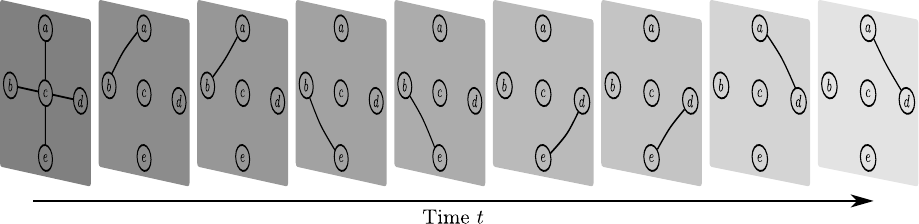}
        \caption{The evolving network changes over time. The connectivity under the temporal constraints is restricted and limits the possibilities for the flow of information.}
        \label{fig:introexamplec}
    \end{subfigure}
    \caption{An example for the $n$-th order temporal H-index. In the temporal network in (a), the $1$-st order temporal H-index of node $c$ is four because $c$ has four outgoing edges at time one, and each of these edges can be extended by four different edges, each with a strictly later time stamp. Nodes $a,b$, and $e$ have a value of two and $d$ of one. In the aggregate static graph (b), all nodes have a H-index value of three.}
    \label{fig:introexample}
\end{figure}

However, recently there has been an increasing focus on dynamically changing networks~\cite{holme2015modern,Michail16,streamgraphs}. 
This trend is motivated by the fact that real-world processes are often inherently dynamic, implying possible causalities that must be taken into account to obtain valid interpretations~\cite{holme2015modern}.
\Cref{fig:introexamplea} shows an example of a small temporal network representing a communication network. The nodes represent persons, and each edge is a bidirectional communication at a specific time shown at the edge. 
\Cref{fig:introexamplec} shows the evolution of the network over time. There is a static graph containing the edges with the corresponding time stamp for each time point.
Note how temporal evolution restricts the possible flow of information.
For example, person $a$ communicates with $b$ at time two and $b$ with $c$ at time four, there is a possibility that information flows from $a$ to $c$ via $b$ but not in the reverse direction due to the chronological order of communications---all communications between $b$ and $c$ are \emph{after} the communications between $a$ and $b$. 
If we only consider the underlying aggregated graph, in which the time stamps are removed, and resulting multi-edges are merged (see \Cref{fig:introexampleb}), there are no restrictions on the possible flow of information.
Therefore, in this work, we focus on analyzing \emph{temporal networks} consisting of a set of nodes and a set of temporal edges endowed with time stamps indicating their time of existence.
Such temporal networks are prime models of real-world interactions like human contact, communication, and social networks~\cite{candia2008uncovering,eckmann2004entropy,holme2004structure,ciaperoni2020relevance,Chaintreau2007}.

\noindent
\textbf{Current work:} We generalize the $n$-th order H-index for temporal networks.
First, we introduce the time-dependent in- and outdegrees of a node $v$ at time $t$, counting the number of incoming or outgoing edges before or after $t$.
On this basis, we define the $n$-th order temporal H-index, respecting temporal reachability and possible implied causality.
We obtain a family of (1) temporal centrality measures and (2) corresponding core-like temporal decompositions.

The centrality measures based on the $n$-th order temporal H-index capture the importance of a node by considering its spatial and temporal position in a neighborhood of depth $n$ in the network. 
We propose two variants---the first measures a node's ability to influence or distribute information, and the second measures how well a node can be influenced or obtain information.
The \emph{1-st order temporal H-index} corresponds to the classic static variant with the critical difference that the temporal restrictions in terms of possible incoming or outgoing information flows are respected.
For example, in \Cref{fig:introexamplea}, node $c$ has four outgoing edges at time $t=1$. Each of the four neighbors also has four outgoing edges at times $t>1$. Hence, any information leaving node $c$ over one of the four edges at the time one can be further distributed via four edges. Therefore, the $1$-st order \emph{outward} temporal H-index is four.
With similar arguments, we see that node $b,c$, and $e$ have a $1$-st order {outward} temporal H-index of two (e.g., $b$ has two outgoing edges to $e$ at times four and five, and $e$ has two outgoing edges later than five). Finally, node $d$ has a value of one.
However, suppose we ignore the temporal information and calculate the conventional static H-index in the underlying aggregated static graph shown in \Cref{fig:introexampleb}. In that case, all nodes obtain the (traditional) H-index value of three, which is uninformative.
Furthermore, if we compute the $1$-st order \emph{inward} temporal H-index \Cref{fig:introexamplea}, we obtain a different ranking of the nodes in 
which $a$ has the smallest index-value and $d$ the highest. Note that for the static H-index, there is no such distinction in undirected networks.
Going from the $1$-st order to higher orders, we can include a deeper neighborhood into the index computation and can identify nodes that
have a strong influence in large parts of the network (or can be influenced by large parts of the network in the inward case).

Moreover, a recursive application of the temporal H-index leads to a core-like decomposition of the network.
Core decompositions are an essential concept in the study of graph properties with many important applications, e.g., (social) network analysis, community detection, or network visualization~\cite{kong2019k,malliaros2020core}.
We define the $(n,k)$-pseudocore in a temporal network $\tg$ as the maximal induced temporal subgraph such that each node has an $n$-th order temporal H-index of at least $k$ with respect to $\tg$.
We again propose two variants based on the in- and outward temporal H-index.
Our evaluation shows that our $n$-th order temporal H-index leads to pseudocores characterized by high temporal reachability.

\noindent
\textbf{Our contributions are:}
\begin{enumerate}[nosep,leftmargin=5mm] %
    \item We introduce the $n$-th order temporal H-index measuring the importance of nodes based on the structure of their neighborhood respecting temporal reachability.
    We propose two variants---the first measures a node's ability to influence or distribute information, and the second measures how well a node can be influenced or obtain information.
    \item Based on the $n$-th order temporal H-index, we introduce a corresponding decomposition of the network into $(n,k)$-pseudocores, describing components with high communication capabilities.
    \item The $n$-th order temporal H-index can be straightforwardly computed by directly implementing its recursive formulation. 
    However, this approach is not scalable, and we propose a highly scalable streaming algorithm operating on the chronologically ordered edge stream using only a single pass over the edges. %
    \item Our evaluation on real-world temporal networks shows that our algorithms are highly efficient even for networks with millions of nodes and tens of millions of temporal edges.
    The $n$-th order temporal H-index effectively identifies central nodes and temporally cohesive subgraphs.
    We demonstrate this in the use case of identifying super-spreaders in epidemic processes in real-world networks.
\end{enumerate}

\smallskip
\noindent
The omitted proofs are provided in \Cref{appendix:proofs}.

\section{Related Work}

There are several thorough surveys and introductions to temporal networks, e.g.,~\cite{holme2015modern,streamgraphs,santoro2011time,Michail16}, centrality measures, e.g., \cite{das2018study,landherr2010critical,rodrigues2019network,Saxena2020}, and core decompositions~\cite{kong2019k,malliaros2020core}.
We focus our discussion on techniques for temporal networks closely related to our work.

\paragraph{Centrality for Temporal Networks}
Several works \cite{nicosia2013graph,kim2012temporal,tang2013applications,tang2010analysing}  examine properties of temporal networks, including various temporal centrality measures, and discuss the necessity of considering temporal dynamics. 
Variants of temporal closeness have been introduced and discussed in~\cite{pan2011path,magnien2015time,santoro2011time,oettershagen2020efficient,crescenziMM20topk}.
Similar to static closeness, the temporal closeness is often defined as the inverse of the sum of optimal temporal distances.
Another path-based centrality measure is betweenness.
\citet{BussMNR20} evaluate the theoretical complexity and practical hardness of computing several variants of temporal betweenness centrality.
\citet{santoro2022onbra} discuss the estimation of temporal betweenness.
\citet{tsalouchidou2019temporal} extend Brandes' algorithm~\cite{brandes2001faster} for distributed computation of betweenness centrality in temporal networks. 
Temporal closeness and betweenness only consider \emph{optimal temporal paths}. Our new centralities, in contrast, consider the general temporal reachability of the node itself and its neighborhood, which is more meaningful in settings where information does not necessarily spread along the shortest paths.
The {Katz centrality}~\cite{katz1953new} measures node importance in terms of the weighted random walks starting (or arriving) at a node. 
The authors of~\cite{katztg,grindrod2011communicability} adapt the walk-based Katz centrality to temporal networks.
\citet{RozenshteinG16} adapt PageRank for temporal networks by using temporal walks. 
\citet{oettershagen2022temporal} generalized the random walk betweenness for temporal networks.
These centrality measures are based on \emph{counting temporal walks}, while our $n$-th order temporal H-index is based on 
\emph{counting neighbors with high temporal reachability through $n$ recursion steps}.
Recent works try to identify influential nodes in spreading processes under the influence maximization model for temporal networks~\cite{erkol2020influence,erkol2022effective}.

\paragraph{Temporal Core Decomposition}
\citet{wu2015core} proposed $(k,h)$-core decompositions for temporal networks, where each node in a $(k,h)$-core has at least $k$ neighbors and at least $h$ temporal edges to each neighbor. The concept can be interpreted as a weighted static core decomposition and does not take temporal dynamics such as reachability into account. 
A recent work by \citet{galimberti2020span} introduced a notion of temporal span-cores where each core is associated with a time span, i.e., a time interval, for which the coreness property holds. 
Let $T$ be the length of the time interval spanned by a temporal graph $\tg$ with edge set $\tge$, then there is a quadratic number of time intervals for which a temporal span-core can exist, and the asymptotic running times of the proposed algorithms are in $\mathcal{O}(T^2\cdot |\tge|)$ which is prohibitive for large networks.
A notable difference to our new approach is that the span-cores decompose a temporal network in its temporal domain such that each vertex may have several distinct span-core values in different time intervals. Whereas our decomposition, similar to the conventional $k$-core, leads to a single core number for each vertex. 

Core-like concepts are also used in temporal community mining.  
\citet{hung2021maximum} define a temporal community as a $(l, k)$-lasting core, which is a $k$-core existing for at least $l$ consecutive time steps and discuss the problem of finding the maximum $(l, k)$-lasting core.
\citet{qin2019mining} consider the densest subgraph problem and define a temporal community as an $(l,\delta)$-maximal dense core which has an average degree of at least $\delta$ in a time interval of length at least $l$. 
\citet{qin2020mining} mine stable communities in temporal networks using the concept of $(\mu,\tau,\epsilon)$-stable cores. Here, a node is in a $(\mu,\tau,\epsilon)$-stable core if it has no less than $\mu$ neighbors that have a similarity of at least $\epsilon$ to the node in at least $\tau$ snapshots of the temporal network.
\citet{li2018persistent} propose as community model the $(\theta,\tau)$-persistent $k$-core variant. It is defined for a time interval $I$ such that a $k$-core of size $\tau$ exists in any subinterval of $I$ with length $\theta$. Computing maximum $(\theta,\tau)$-persistent $k$-cores is $\mathsf{NP}$-hard. 

Our $n$-th order temporal H-index leads to a \emph{decomposition that has only a single parameter and allows efficient computation} (see \Cref{tab:overview}).
Core decompositions are valuable tools, e.g., in social network analysis~\cite{shin2016corescope}, community detection~\cite{rossetti2018community}, and network visualization~\cite{alvarez2006k}.  
The previously mentioned approaches do not take temporal reachability explicitly into account, whereas \emph{ours supports this naturally via the $n$-th order temporal H-index}.

\section{Preliminaries}
We introduce basic definitions and our notation. 
We refer to the natural numbers without zero by $\mathbb{N}$, and define $\mathbb{N}_0 = \mathbb{N} \cup \{0\}$.
A (static) \emph{graph} $G=(V, E)$ consists of a finite set of nodes $V$ and a finite set $E\subseteq\{\{u,v\}\subseteq V\mid u\neq v\}$ of undirected edges.
A node $v\in V$ is \emph{incident} to $e\in E$ if $v\in e$. 
The degree $\delta(v)$ of a node $v\in V$ is the number of edges incident to $v$.

\subsection{Temporal Networks}
An \emph{undirected temporal network} $\tg=(V, \tge)$ consists of a finite set of nodes $V$ and a finite set $\tge$ of undirected \emph{temporal edges} $e=(\{u,v\},t,\lambda)$ with $u$ and $v$ in $V$, $u\neq v$, \emph{availability time} (or \emph{time stamp}) $t \in \mathbb{N}$ and \emph{transition time} (or \emph{traversal time}) $\lambda \in \mathbb{N}$. The availability time specifies when the transition from $u$ to $v$ or $v$ to $u$ via the edge is possible. 
In a \emph{directed} temporal network, a temporal edge has the form $(u,v,t,\lambda)$ and can be traversed from $u$ to $v$ only. 
For convenience, we primarily consider directed temporal networks in the following, which allow to model undirected edges by means of two symmetric directed edges.
The transition time of an edge is the time required to traverse the edge, e.g., in a temporal transportation network the time a vehicle needs between two stops, or in a human contact network the time to give information from one person to another person.
Notice that it is reasonable and common to assume a strictly positive transition time as information in standard models cannot be transferred instantaneously.
The in- and out-degrees of a node $v$ count the temporal edges arriving or leaving, resp., at $v$.
We use $\maxdeg^-(\tg)$ to denote the maximal in-degree and $\maxdeg^+(\tg)$ for the maximal out-degree of $\tg$. We write $\maxdeg^-$ or $\maxdeg^+$ for short if $\tg$ is clear from the context.

Given a (directed or undirected) temporal network $\tg$, removing all time stamps and traversal times, and merging resulting parallel edges, we obtain the (directed or undirected) aggregated graph $A(\tg)=(V,E_s)$ with $E_s=\{(u,v)\mid(u,v,t,\lambda)\in\tge\}$.

Given a temporal network $\tg=(V,\tge)$, it is common to restrict research questions to a given time interval
$I=[a,b]$, such that only the temporal subgraph $\tg'=(V,\tge')$ with $\tge'=\{(u,v,t,\lambda)\in \tge \mid t\geq a \text{ and } t+\lambda \leq b\}$ needs to be considered.
We do not include restrictive intervals in the following definitions for better readability. However, our definitions and algorithms can be easily adapted to consider only temporal edges starting and arriving within the interval. 
    A \emph{temporal walk} of length $\ell$ in a temporal network $\tg$ is an alternating sequence $(v_1, e_1,\ldots,e_\ell,v_{\ell+1})$ of nodes and temporal edges connecting consecutive nodes, such that
    for $1\leq i < \ell$, $e_i=(v_i,v_{i+1},t_i,\lambda_i)\in \tge$ and $e_{i+1}=(v_{i+1},v_{i+2},t_{i+i},\lambda_{i+1})\in \tge$, and the time constraint $t_i+\lambda_{i}\leq t_{i+1}$ holds. 
Let $\Delta(\tg)$ be the \emph{temporal diameter} of $\tg$, i.e., the length of the longest (in terms of number of edges) temporal walk in $\tg$.

\subsection{The H-Index and k-Core Decomposition}

A \emph{$k$-core} of a graph $G$ is a maximal subgraph $G_k$ of $G$, such that every node in $G_k$ has at least $k$ neighbors in $G_k$.
A node $u$ has \emph{core number} $c(u)=k$ if $u$ belongs to a $k$-core but not the $k+1$-core of the graph~\cite{malliaros2020core}.
The core numbers of all nodes can be computed in $\mathcal{O}(|V|+|E|)$ time~\cite{batagelj2003m}.

\citet{lu2016h} showed a relationship between degree centrality, H-index, and core numbers. To this end, they introduced the $\mathcal{H}$ operator, which essentially represents the calculation of the H-index. 
    Let $\mathcal{S}$ be the set of finite multisets of integers.
    The function $\hop\colon \mathcal{S} \rightarrow \mathbb{N}_0$ returns for a finite multiset of integers $S\subseteq\{\!\!\{s\mid s\in\mathbb{N}_0\}\!\!\}$ 
    the maximum integer $i$ such that there are at least $i$ elements $s$ in $S$ with $s\geq i$.  
    We define $\hop(\emptyset)=0$.

\begin{figure*}
    \begin{subfigure}[b]{.34\linewidth}
        \centering
        \includegraphics[width=0.8\linewidth]{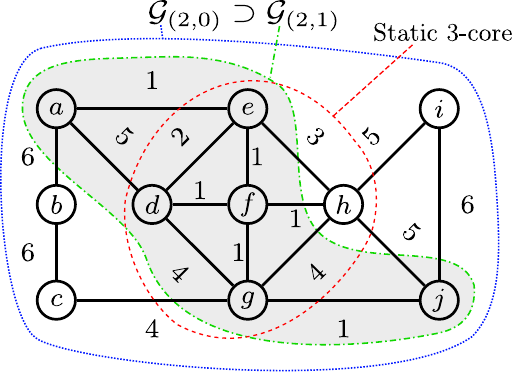}
        \caption{Temporal network $\tg$ with the availability times at the edges ($\lambda=1$ for all edges).
            The out-pseudocore-decomposition for $n=2$ is shown.
        }
        \label{fig:example1a}
    \end{subfigure}\hspace{2mm}\hfill%
    \begin{subfigure}[b]{.64\linewidth}
        \centering
        \begin{tikzpicture}[scale=0.7]
            \begin{scope}
                \node[font=\scriptsize] (d0) at (16.7,5.4) {Depth:};
                \draw[black!50!white,dashed] (1.5,5) -- (16.5,5) node[right,black,font=\scriptsize] (d0) {$0$};
                \draw[black!50!white,dashed] (1.5,4) -- (16.5,4) node[right,black,font=\scriptsize] (d1) {$1$};
                \draw[black!50!white,dashed] (1.5,3) -- (16.5,3) node[right,black,font=\scriptsize] (d2) {$2$};
                \draw[black!50!white,dashed] (1.5,2) -- (16.5,2) node[right,black,font=\scriptsize] (d3) {$3$};
                \draw[black!50!white,dashed] (1.5,1) -- (16.5,1) node[right,black,font=\scriptsize] (d4) {$4$};
                \draw[black!50!white,dashed] (1.5,0) -- (16.5,0) node[right,black,font=\scriptsize] (d5) {$5$};
                \node[treevertex,color=black,fill=white] (f00) at (8.75,5) {$f,\!0$}; 
                \node[treevertex,color=black,fill=white] (d21) at (3,4) {$d,\!2$};
                \node[treevertex,color=black,fill=white] (e21) at (7,4) {$e,\!2$};
                \node[treevertex,color=black,fill=white] (h21) at (10.5,4) {$h,\!2$};
                \node[treevertex,color=black,fill=white] (g21) at (14,4) {$g,\!2$};
                \node[treevertex,color=black,fill=white] (e32) at (3,3) {$e,\!3$};
                \node[treevertex,color=black,fill=white] (a62) at (4,3) {$a,\!6$};
                \node[treevertex,color=black,fill=white] (g52) at (2,3) {$g,\!5$};
                \node[treevertex,color=black,fill=white] (d32) at (6,3) {$d,\!3$};
                \node[treevertex,color=black,fill=white] (h52-2) at (8,3) {$h,\!4$};
                \node[treevertex,color=black,fill=white] (e42) at (9,3) {$e,\!4$};
                \node[treevertex,color=black,fill=white] (i62) at (10,3) {$i,\!6$};
                \node[treevertex,color=black,fill=white] (j62) at (11,3) {$j,\!6$};
                \node[treevertex,color=black,fill=white] (g52-3) at (12,3) {$g,\!5$};
                \node[treevertex,color=black,fill=white] (c52) at (13,3) {$c,\!5$};
                \node[treevertex,color=black,fill=white] (d52) at (14,3) {$d,\!5$};
                \node[treevertex,color=black,fill=white] (h52) at (15,3) {$h,\!5$};
                \node[treevertex,color=black,fill=white] (h43) at (3,2) {$h,\!4$};
                \node[treevertex,color=black,fill=white] (b73) at (4,2) {$b,\!7$};
                \node[treevertex,color=black,fill=white] (a63) at (6,2) {$a,\!6$};
                \node[treevertex,color=black,fill=white] (g53) at (5,2) {$g,\!5$};
                \node[treevertex,color=black,fill=white] (j73) at (10,2) {$j,\!7$};
                \node[treevertex,color=black,fill=white] (i73) at (11,2) {$i,\!7$};
                \node[treevertex,color=black,fill=white] (b73-2) at (13,2) {$b,\!7$};
                \node[treevertex,color=black,fill=white] (a63-2) at (14,2) {$a,\!6$};
                \node[treevertex,color=black,fill=white] (i63) at (15,2) {$i,\!6$};
                \node[treevertex,color=black,fill=white] (j63) at (16,2) {$j,\!6$};
                \node[treevertex,color=black,fill=white] (i64) at (2,1) {$i,\!6$};
                \node[treevertex,color=black,fill=white] (j64) at (3,1) {$j,\!6$};
                \node[treevertex,color=black,fill=white] (b74) at (6,1) {$b,\!7$};
                \node[treevertex,color=black,fill=white] (b74-2) at (14,1) {$b,\!7$};
                \node[treevertex,color=black,fill=white] (j74) at (15,1) {$j,\!7$};
                \node[treevertex,color=black,fill=white] (i74) at (16,1) {$i,\!7$};
                \node[treevertex,color=black,fill=white] (j75) at (2,0) {$j,\!7$};
                \node[treevertex,color=black,fill=white] (i75) at (3,0) {$i,\!7$};
                
                \node[treevertex,color=black,fill=white] (j84) at (8,2) {$j,\!6$};
                \node[treevertex,color=black,fill=white] (i84) at (7,2) {$i,\!6$};
                \node[treevertex,color=black,fill=white] (j85) at (8,1) {$i,\!7$};
                \node[treevertex,color=black,fill=white] (i85) at (7,1) {$j,\!7$};
                
                \node[treevertex,color=black,fill=white] (g54) at (4,1) {$g,\!5$};
                \node[treevertex,color=black,fill=white] (g65) at (9,2) {$g,\!5$};
            \end{scope}
            
            \begin{scope}[>={Stealth[black]},
                every node/.style={circle},
                ]
                \path [edge] (f00) edge (d21);
                \path [edge] (f00) edge (e21);
                \path [edge] (f00) edge (h21);
                \path [edge] (f00) edge (g21);
                \path [edge] (d21) edge (e32);
                \path [edge] (d21) edge (a62);
                \path [edge] (d21) edge (g52);
                \path [edge] (e21) edge (d32);
                \path [edge] (e21) edge (h52-2);
                \path [edge] (h21) edge (e42);
                \path [edge] (h21) edge (i62);
                \path [edge] (h21) edge (j62);
                \path [edge] (h21) edge (g52-3);
                \path [edge] (g21) edge (c52);
                \path [edge] (g21) edge (d52);
                \path [edge] (g21) edge (h52);
                \path [edge] (e32) edge (h43);
                \path [edge] (h43) edge (i64);
                \path [edge] (h43) edge (j64);
                \path [edge] (i64) edge (j75);
                \path [edge] (j64) edge (i75);
                \path [edge] (a62) edge (b73);
                \path [edge] (d32) edge (a63);
                \path [edge] (a63) edge (b74);
                \path [edge] (d32) edge (g53);
                \path [edge] (i62) edge (j73);
                \path [edge] (j62) edge (i73);
                \path [edge] (c52) edge (b73-2);
                \path [edge] (d52) edge (a63-2);
                \path [edge] (a63-2) edge (b74-2);
                \path [edge] (h52) edge (i63);
                \path [edge] (h52) edge (j63);
                \path [edge] (i63) edge (j74);
                \path [edge] (j63) edge (i74);
                \path [edge] (h52-2) edge (i84);
                \path [edge] (h52-2) edge (j84);
                \path [edge] (i84) edge (i85);
                \path [edge] (j84) edge (j85);
                \path [edge] (h43) edge (g54);
                \path [edge] (h52-2) edge (g65);
            \end{scope}
        \end{tikzpicture}
        \caption{The reachability tree $\Gamma^+(f)$ for vertex $f$ in the temporal network shown in (a).}
        \label{fig:example1b}
    \end{subfigure}
    \caption{Toy example for the $n$-th order temporal H-index. \Cref{table:example1} shows the corresponding values for all vertices.}
    \label{fig:example1}
\end{figure*}
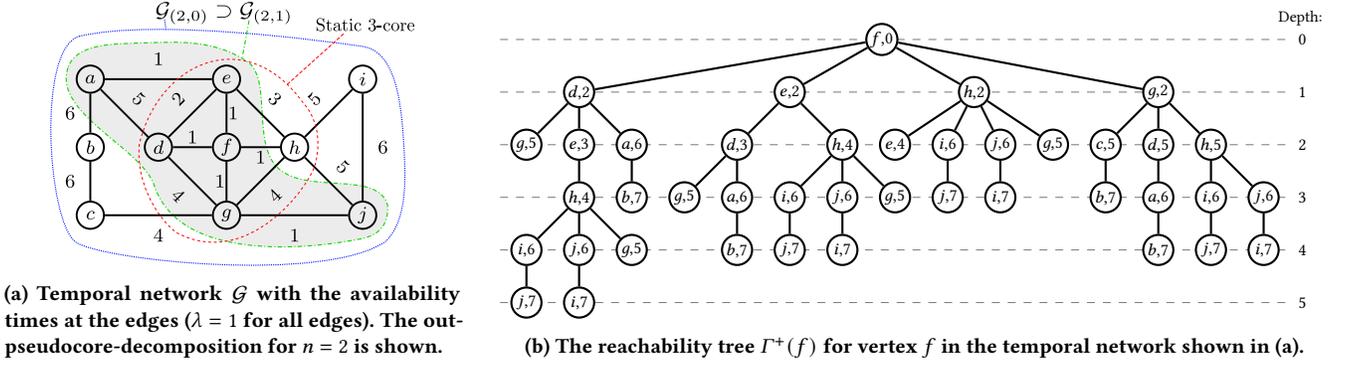

Using the $\mathcal{H}$ operator, the H-index of a node $u$ is defined as $H(u)=\hop(\{\!\!\{\delta(v)\mid v \in V \text{ and $v$ is neighbor of $u$}\}\!\!\})$.
The $n$-th order H-index $\sh{u}{n}$ of a node $u$ in a static graph is defined recursively~\cite{lu2016h}.
Let $\sh{v}{0}=\delta(v)$ the degree of node $v$, then
\[
\sh{u}{n}=\hop\left(\{\!\!\{\sh{v}{n-1}\mid v \in V \text{ and $v$ is neighbor of $u$}\}\!\!\}\right).
\]
For $n=0$ and $n=1$, the value of $\sh{u}{n}$ corresponds to the degree and H-index of $u$, respectively.
For increasing $n$, it converges to the \emph{core number} of $u$~\cite{lu2016h}.

\section{The n-th Order Temporal H-Index}
We now define the $n$-th order H-index for temporal networks.
To this end, we start by generalizing the degree of a node to incorporate the temporal aspects of the local network topology.

\begin{definition}\label{def:degrees}
    Given a temporal network $\tg=(V,\tge)$, we define the \emph{time-dependent
    in-neighborhood} of a node $u$ at time $t$ as 
    ${N^-}(u,t)=\{(v,u,t_e,\lambda_e)\in \tge \mid t_e+\lambda_e\leq t \}$.
    Similarly, the \emph{time-dependent out-neighborhood} is
    ${N^+}(u,t)=\{(u,v,t_e,\lambda_e)\in \tge \mid t_e\geq t \}$.
    Finally, we define $\tdegin{u}{t}=|N^-(u,t)|$ and $\tdegout{u}{t}=|N^+(u,t)|$ the time-dependent in- and out-degree, respectively.
\end{definition}
The time-dependent out-degree of $u$ reflects the number of ways to extend a temporal walk from $u$ at or after time $t$ by an edge. The time-dependent in-degree counts the number of edges via which a temporal walk can reach $u$ no later than at time $t$. 
Furthermore, we define sets of nodes with time stamps that allow extending walks.
\begin{definition}\label{def:nodetimes}
    We define the multiset $\mathcal{N}^{+}(v,t)$ that contains all pairs of nodes and times $(w,t_w)$ such that there is a temporal edge $(v,w, t', \lambda)\in \tge$ 
    with arrival time $t_w=t'+\lambda$ and $t'\geq t$.
    Analogously, we define the multiset $\mathcal{N}^{-}(v,t)$ that contains all pairs of nodes and starting times $(w,t_w)$ such that there exists a temporal edge $(w,v, t_w, \lambda)\in \tge$ 
    with $t_w+\lambda\leq t$ . 
\end{definition}

Using the \Cref{def:degrees,def:nodetimes} together with the $\mathcal{H}$ operator, 
we define the $n$-th order temporal H-index.
\begin{definition}\label{def:nhix}
    Let $(\square,\star)\in \{(0,+), (\infty,-)\}$.
    The \emph{$n$-th order temporal H-index} of a node $v\in V$ is defined as
    $\hfh{v,\star}{n}=\hf{v}{\square,\star}{n}$ with
    \[\hf{v}{t,\star}{n}=\hop\left(\left\{\!\!\left\{\hf{w}{t_w,\star}{n-1} 
    \;\middle|\; 
    (w,t_w)\in \mathcal{N}^{\star}(v,t) \right\}\!\!\right\}\right).\]
    We define $\hf{v}{t,\star}{0}=\delta^{\star}(v,t)$. %
\end{definition}
We call $\hfh{v,+}{n}$ the \emph{outward} {$n$-th order temporal H-index} and
$\hfh{v,-}{n}$ the \emph{inward} {$n$-th order temporal H-index}.
Note that also for undirected temporal networks typically $\hfh{v,+}{n} \neq \hfh{v,-}{n}$ due to the direction implied by the temporal information.  

The $n$-th order temporal H-index equals the (in-/out-) degree centrality in the case of $n=0$.
For $n=1$, we obtain time-directed variants of the classical H-index.
For larger $n$, intuitively, a node $v$ has a high index value if it is 
in a temporally well-connected position in the network.
More precisely, for the outward variant, $v$ can reach many other nodes with temporal walks of length $n+1$,
and this recursively holds for the reachable nodes (with reduced walk length).
Therefore, nodes with high outward temporal H-index can strongly influence their neighborhood, e.g., spread information easily, and are connected to similarly influential nodes. 
Analogously, a node $v$ has a high inward index value if the node can be reached by many other nodes with temporal walks of length $n+1$, and this recursively holds for the nodes reaching $v$ (with reduced walk length).
Hence, nodes with high inward temporal H-index can easily obtain information, i.e., be influenced, and are also connected to nodes with a similar ability.
For both variants, the influence of temporal walk counts is controlled by the iterated application of the $\hop$ operator based on the temporal structure of the neighborhood. 
We determine the exact lower bounds of the number of reachable nodes in \Cref{theorem:numdescendants}.
\Cref{fig:example1a} gives an example of a temporal network, and \Cref{table:example1} shows the outward temporal H-index values of the nodes for $n\leq 4$. 

\Cref{table:example2} shows the values of the static $n$-th order H-index for the underlying static graph ${A}(\tg)$, ignoring the temporal information.
The values of the H-index $(n=1)$ and $k$-core numbers $(n=2)$ differ for all nodes. Likewise, for $n=1$, the \emph{rankings} of the nodes disagree, as in the static case, the nodes are only ranked in three groups (i.e., nodes with value $2,3$, or $4$) and in the temporal case in four (i.e., nodes with values $0,1,2$, or $3$, respectively).
For $n=2$, the decompositions differ in the nodes $a$, $h$, and $j$. As \Cref{fig:example1a} shows, the most inner core $\tg_{(n,1)}$ (for $n=2$, highlighted in green) contains both $a$ and $j$, but not $h$.
The static $k$-core decomposition decomposes the graph into two groups of nodes with core numbers two and three. The most inner core, i.e., induced by nodes with the core number $k=3$, contains $h$ but not $a$ and $j$ (highlighted in red).

\begin{table}[t]
    \caption{The values of the static and temporal $H$-index for the example shown in \Cref{fig:example1}.}
    \begin{subtable}{1\linewidth}
        \centering
        \caption{The values $\hfh{v,+}{n}$ of the outward $n$-th order temporal $H$-index. }
        \label{table:example1}
        \resizebox{1\linewidth}{!}{\renewcommand{\arraystretch}{0.8}\setlength{\tabcolsep}{2.8mm}
            \begin{tabular}{l@{\hspace{8mm}}cccccccccc}\toprule
                &\multicolumn{10}{c}{$n$-th Order Temporal $H$-index $\hfh{v,+}{n}$}\\
                \cmidrule{2-11}
                $n$ & $a$ & $b$ & $c$ & $d$ & $e$ & $f$ & $g$ & $h$ & $i$ & $j$ \\ \midrule
                $0$ (degree)&  3  &  2  &  2  &  4  &  4  &  4  &  5  &  5  &  2  &  3  \\ 
                $1$         &  1  &  0  &  0  &  1  &  2  &  3  &  2  &  1  &  0  &  1  \\ 
                $2$         &  1  &  0  &  0  &  1  &  1  &  1  &  1  &  0  &  0  &  1  \\
                $3$         &  1  &  0  &  0  &  1  &  0  &  1  &  0  &  0  &  0  &  1  \\
                $4$         &  0  &  0  &  0  &  0  &  0  &  1  &  0  &  0  &  0  &  0  \\
                \bottomrule
            \end{tabular}
        }
    \end{subtable}
    \begin{subtable}{1\linewidth}
        \centering\vspace{1mm}
        \caption{The values $\sh{v}{n}$ of the $n$-th order static $H$-index computed in the underlying static graph. }
        \label{table:example2}
        \resizebox{1\linewidth}{!}{\renewcommand{\arraystretch}{0.8}\setlength{\tabcolsep}{2.8mm}
            \begin{tabular}{l@{\hspace{8mm}}cccccccccc}\toprule
                &\multicolumn{10}{c}{$n$-th Order Static $H$-index $\sh{v}{n}$}\\
                \cmidrule{2-11}
                $n$ & $a$ & $b$ & $c$ & $d$ & $e$ & $f$ & $g$ & $h$ & $i$ & $j$ \\ \midrule
                $0$ (degree)    &  3  &  2  &  2  &  4  &  4  &  4  &  5  &  5  &  2  &  3  \\ 
                $1$ ($H$-index) &  2  &  2  &  2  &  3  &  3  &  4  &  3  &  3  &  2  &  2  \\ 
                $2$ ($k$-core)  &  2  &  2  &  2  &  3  &  3  &  3  &  3  &  3  &  2  &  2  \\
                \bottomrule
            \end{tabular}
        }
    \end{subtable}	
\end{table}

\subsection{Properties}\label{sec:properties} %
In the following, we investigate the properties of the $n$-th order temporal H-index. %
To this end, we first define a \emph{reachability tree} that represents all outgoing (or incoming) temporal walks from a node $v\in V$.
The reachability tree is not a data structure used in our algorithms but a valuable tool for illustrating and discussing the $n$-th order temporal H-index. 
All results in this section hold for the outward and inward variants of the $n$-th order temporal H-index.

\begin{definition}\label{def:reachtree}
    Let $\star\in\{-,+\}$, %
    $\tg=(V,\tge)$ a temporal network, $v\in V$, and $t\in\mathbb{N}_0$. 
    The tree $\Gamma^\star(v,t)=(U, E, r)$ with $U \subseteq \{(u,t')\mid u\in V, t'\in \mathbb{N}_0\}$ is defined recursively as the tree with root $r=(v,t)$ and a child $c_i$ for each $(w_i,t_i)\in \mathcal{N}^\star(v,t)$, where $c_i$ is the root of the tree $\Gamma^\star(w_i,t_i)$.
For $\star=+$, we define $\Gamma^+(v)=\Gamma^+(v,0)$, and for $\star=-$, $\Gamma^-(v)=\Gamma^-(v,\infty)$.

\end{definition}

Let $d(u)$ be is the depth of $u \in V(\Gamma^\star(v))$.
The tree $\Gamma^+(v)$ contains a node $u=(v,t)$ for each node that we can reach at time $t$ in a temporal walk starting at node $v$ using $d(u)$ consecutive temporal edges.
For example, \Cref{fig:example1b} shows $\Gamma^+(f)$ for the node $f$ of the temporal network in \Cref{fig:example1a}.
Similarly, $\Gamma^-(v)$ contains a node $u=(v,t)$ for each node that can reach $v$ with a temporal walk starting at a time $t$ using $d(u)$ consecutive temporal edges.
\begin{definition}\label{def:treephi}
    Let $\star\in\{-,+\}$, $\Gamma^\star(v)=(U, E,r)$, $u\in U$, and $C(u)\subseteq U$ be the children of $u$.
    For $n\in\mathbb{N}_0$, we define $\phi_n \colon U\rightarrow \mathbb{N}_0$ as
    \begin{equation}
        \phi_n(u)=
        \begin{cases}
            \begin{aligned}
                &\hi{\{\!\!\{ \phi_n(c) \mid \text{$c\in C(u)$}\}\!\!\}}, &\text{ if $d(u)< n$}\\
                 &|C(u)|, &\text{ if $d(u)=n$}\\
                &0, &\text{ otherwise.}
            \end{aligned}
        \end{cases}
    \end{equation}
\end{definition}
The usefulness of $\Gamma^\star(v)$ together with $\phi_n$ is given by the following result.
\begin{lemma}\label{lemma:phihf}
    Let $\star\in\{-,+\}$, $\tg=(V,\tge)$ be a temporal network, $v\in V$, $\Gamma^\star(v)=(U,E,r)$ be the reachability tree of $v$, and
    $n\in\mathbb{N}_0$. Then, $\phi_n(u=(w,t))=\hf{w}{t,\star}{n-d(u)}$ for each $u\in U$ with $d(u)\leq n$.
\end{lemma}

Using $\Gamma^\star(v)$, we can state a lower bound based on $\hfh{v,\star}{n}$ of the number of different temporal walks of length at most $n$ starting from node $v$. 
\begin{theorem}\label{theorem:numdescendants}
    Let $\star\in\{-,+\}$, $\tg=(V,\tge)$ be a temporal network, $v\in V$, $n\in \mathbb{N}$, and $\hfh{v,\star}{n}=k$.
    Then, there are at least $({k^{(n+2)} - k})/({k-1})$
    descendants $u\in U$ of the root $r$ in $\Gamma^\star(v)$ with $d(u)\leq n$.  
\end{theorem}

Each descendant of $r$ in $\Gamma^\star(v)$ corresponds to one temporal walk leaving ($\star=+$) or arriving ($\star=-$) at $v$ in $\tg$.
Hence, the $n$-th order temporal H-index, on the one hand, gives a lower bound of outgoing/incoming temporal walks of $v$. 
On the other hand, for each visited node a similar lower bound holds, i.e., each descendant $u$ of $r$ with depth $d_u$ has at least $\sum_{i=1}^{n-d_u+1} k^i=\frac{k^{n-d_u+2}-k}{k-1}$ descendants.

In static networks, the static $n$-th order H-index of a node $v\in V$ converges to the $k$-core value of the node $v$ for sufficiently large $n$~\cite{lu2016h}. This does not hold for the $n$-th order temporal H-index.

\begin{theorem}\label{theorem:largentozero}
    Let $\star\in\{-,+\}$, $\tg=(V,\tge)$ and $v\in V$. It holds that %
    $\hfh{v,\star}{n}=0$ for all $n>\Delta(\tg)$.
\end{theorem}

Therefore, neither the outward nor the inward $n$-th order temporal H-index converges to the $k$-core composition of the network.

\subsection{Temporal Pseudocore Decomposition}
The two variants of the $n$-th order temporal H-index provide two natural ways of decomposing the temporal network.
\begin{definition}\label{def:nkcore}
    Let $\star\in\{-,+\}$, $\tg=(V,\tge)$ be  a temporal network, and $k,n\in\mathbb{N}$.
    The \emph{temporal $(n,k)$-pseudocore} of $\tg$ is a maximal induced temporal subgraph $\tg_{(n,k)}$ of $\tg$ such that for all $v\in V(\tg_{(n,k)})$
    the $n$-th order temporal H-index $\hfh{v,\star}{n}\geq k$ in $\tg$.
\end{definition}
We distinguish \emph{in}- and \emph{out}-pseudocores for $\star=-$ and $\star=+$, resp.
The following observations and results hold for both variants.
For a node $v$ in a $(n,k)$-pseudocore $\tg_{(n,k)}$, the inequality $\hfh{v,\star}{n}\geq k$ does not hold necessarily with respect to $\tg_{(n,k)}$.
However, each $(n,k)$-pseudocore is a subset of nodes that implies a temporal subgraph containing nodes with similar temporal activity and importance in the network $\tg$. 
In~\Cref{sec:applications}, we show that the pseudocore can be used to identify temporally well-connected subgraphs.

\begin{definition}
    For a fixed $n\in\mathbb{N}$, the \emph{temporal pseudo-degeneracy} $\tau_n$ of a temporal network $\tg$ is defined as the maximum 
    $k$ for which $\tg$ contains a non-empty $(n,k)$-pseudocore.
    Similarly, for a fixed $k\in\mathbb{N}$, we define the \emph{order pseudo-degeneracy} $\eta_k$ as the maximum $n$ for which $\tg$ contains a non-empty $(n,k)$-pseudocore. %
\end{definition}

We show the following containment properties. %

\begin{theorem}\label{theorem:containment}
    Let $\tg=(V,\tge)$ be  a temporal graph and $k,n\in\mathbb{N}$.
    For $n\in\mathbb{N}$ fixed, it holds
        $\tg_{(n,0)}\supseteq \tg_{(n,1)}\supseteq\ldots\supseteq\tg_{(n,\tau_n-1)}\supseteq\tg_{(n,\tau_n)}$,
    and for $k\in\mathbb{N}$ fixed, 
        $\tg_{(0,k)}\supseteq \tg_{(1,k)}\supseteq\ldots\supseteq\tg_{(\eta_k-1,k)}\supseteq\tg_{(\eta_k,k)}$.
\end{theorem}

\Cref{fig:example1a} shows an example of the out-pseudocore decomposition for $n=2$, and
\Cref{table:example2} shows the core numbers of the (conventional) $k$-core decomposition of the underlying aggregated graph $A(\tg)$ that ignores the temporal information.
Note the differences for nodes $a$, $j$, and $h$ between the rankings of the static $k$-core and the temporal out-pseudocore (note that we compare the relative rankings and not the core values, which are not directly comparable) resulting from the conceptual differences of the out-pseudocore and the conventional $k$-core.

\subsection{Computation}
We discuss two algorithms with different properties and running times for computing the $n$-th order temporal H-index.
Both algorithms can be used to compute the inward and outward variants. 
The first algorithm \textsc{Recurs} is a straightforward implementation of the recursive formulation.
We additionally use memoization to avoid redundant computations.
However, the running time of \textsc{Recurs} still is unsatisfactory.
Therefore, we introduce a highly efficient one-pass streaming algorithm based on the \emph{edge stream representation} of temporal networks, which represents the temporal network as a chronologically ordered list of the temporal edges (ties are broken arbitrarily). 
The edge stream representation is commonly used for temporal graph streaming algorithms, e.g.,~\cite{wu2014path,mutzel2019enumeration,oettershagen2022temporal}. 
Our streaming algorithm leads to significantly improved running times compared to the recursive algorithm (see \Cref{sec:efficiency}).
It only supports temporal networks with equal transition times for all temporal edges. However, this is a common property for many real-world temporal networks, e.g., face-to-face contact or email networks. 

\Cref{tab:overview} shows an overview of the algorithms and their properties.
We provide the pseudocode and a discussion of the running time of \algr in \Cref{appendix:computation}.
Both algorithms can be easily adapted to only consider edges in a given interval $I$. Alternatively, we can remove all edges not starting or arriving in $I$ in a preprocessing step in $\mathcal{O}(|\tge|)$ time. 
Moreover, we assume that no isolated nodes exist, such that $|V|\leq 2|\tge|$. 
Because the degree of an isolated node $v$ is zero, it follows $\hfh{v}{i}=0$ for all $0\leq i \leq n$.
Hence, we can safely remove all isolated nodes in a preprocessing step in $\mathcal{O}(|V|)$ time.

\Cref{tab:overview2} in the appendix gives an overview of the complexities of related temporal centrality measures and core decompositions, showing the competitiveness of our streaming algorithm.

\begin{table}[t]
    \centering
    \caption{Overview of the algorithms for computing the $n$-th order temporal H-index and their properties ($\star=-$ for the inward variant and $\star=+$ for the outward variant.)}
    \label{tab:overview}
    \resizebox{\linewidth}{!}{\renewcommand{\arraystretch}{1.1}\setlength{\tabcolsep}{3pt}
        \begin{tabular}{ccccc}\toprule
            \textbf{Algorithm}   & \textbf{Running Time} & \textbf{Space}  & \textbf{Edge Trans. Times} & \textbf{Results for} $\forall i\in[n]$\\ \midrule
            \textsc{Recurs}  & $\mathcal{O}(|V| n (\maxdeg^\star)^2)$  & $\mathcal{O}(|V| n \maxdeg^\star)$  & individual & \xmark \\
            \textsc{Stream}  & $\mathcal{O}(|\tge| n  \maxdeg^\star)$  & $\mathcal{O}(|V| n \maxdeg^\star)$  & uniform & \cmark \\ 
            \bottomrule
        \end{tabular}
    }	
\end{table}

\medskip
In the following, we assume that $\lambda=1$ for all $e\in \tge$.
\Cref{alg:streaming} shows the pseudocode of our streaming algorithm. It needs only a single pass over the edges in reverse chronological order, where temporal edges with equal availability times are processed in an arbitrary order.
The algorithm computes $\hfh{v,+}{i}$ \emph{for all} $0\leq i \leq n$ and $v\in V$ (note that \Cref{alg:topdown} only computes $\hfh{v,\star}{n}$ for a fixed $n$). 
In \Cref{appendix:revcomp}, we show how to transform the input to use \Cref{alg:streaming} to compute $\hfh{v,-}{i}$ {for all} $0\leq i \leq n$ and $v\in V$.

A key idea of the algorithm is to keep a single counter for the time-dependent degree for each node.
To this end, we use two arrays $lt$ and $deg$; 
the former stores the last time $deg$ was updated, and $deg$ contains the current time-dependent degree of node $u$.
Furthermore, the algorithm manages a list $\pi[v][i]$ of pairs $(t',x)$ of $i$-th order $H$-indices for $0\leq i \leq n+1$ and for all $v\in V$.
Here, the time $t'$ is stored to exclude values during the computation of the $\hop_t$ operator depending on some time $t$ (see line~\ref{alg:streaming:htfilter}).
\begin{theorem}\label{theorem:algstream}
    \Cref{alg:streaming} computes $\hfh{v,+}{i}$ for all $0\leq i \leq n$ and $v\in V$ in time
    $\mathcal{O}(|\tge|\cdot n \cdot \maxdeg^+)$ with space $\mathcal{O}(|V|\cdot n\cdot \maxdeg^+)$.
\end{theorem}

\begin{algorithm}[h]\small
    \label[algorithm]{alg:streaming}
    \caption{Streaming algorithm \textsc{Stream}}
    \Input{Temporal graph $\tg=(V,\tge)$ as ordered edge stream, $n\in\mathbb{N}$ 
}   
    \Output{$\hfh{v,+}{i}$ for all $0\leq i\leq n$ and $v\in V$}

    \BlankLine
    \SetKwProg{Fn}{Function}{:}{}
    \SetKwFunction{fhit}{$\mathcal{H}_t$} \tcp{$\hit{t}{L}$ computes $\hi{L}$ dependent on time $t$}
    \Fn{\fhit{$L$}}{\label{alg:streaming:htfilter}
        $L_t \gets$ $\{\!\!\{x \mid (t',x)\in L$ with $t'> t \}\!\!\}$\;
        \Return $\hi{L_t}$\;
    }
    \BlankLine  
    
    Initialize $lt[v]=\infty$ and $deg[v]=0$ for all $v \in V$\;
    Initialize list $\pi[v][i]$ for each $v\in V$ and $0\leq i \leq n+1$\;

    \ForEach{$e=(u,v,t)\in\tge$ in reverse chronological order}{\label{alg:streaming:forloop} 
        \If {$lt[u] > t$} {
            $lt[u] \gets t$\;
            $deg[u] \gets \pi[u][0].length$\;
        }
        \If {$lt[v] > t$} {
            $lt[v] \gets t$\;
            $deg[v] \gets \pi[v][0].length$\;
        }
        
        $\pi[u][0].append((t, 1))$\;\label{alg:streaming:updatepione}
        
        \lIf {$n = 0$}{{continue}
        }
        
        $\pi[u][1].append((e.t, deg[v]))$\;\label{alg:streaming:updatepideg}
        
        \lFor {$1\leq j \leq n$} {\label{alg:streaming:forloopn}
            $\pi[u][j + 1].append((t, \hit{t}{\pi[v][j]}))$
        }
        
    }
    
    \BlankLine
    Initialize $h^{(i)}_{v,+}=0$ for all $v\in V$ and $0\leq i \leq n$\;\label{alg:streaming:forloopfinalstart} %
    \For {$v\in V$} {
        $h^{(0)}_{v,+}\gets \pi[v][0].length$\;\label{alg:streaming:nzero}
        \lFor {$1\leq i \leq n$} {
            $h^{(i)}_{v,+}\gets \hit{0}{\pi[v][i]}$
        }
    }\label{alg:streaming:forloopfinalend}
    \Return $h^{(i)}_{v,+}$ for all $v\in V$ and $0\leq i \leq n$\;
\end{algorithm}

\section{Experimental Evaluation}\label{sec:experiments}
We experimentally evaluate the efficiency of our algorithms and the efficacy of the $n$-th order temporal H-index.

\noindent
\textbf{Data Sets: }
We use a wide range of real-world temporal network data sets from various online and offline domains for our evaluation.
\Cref{table:datasets_stats2} shows the properties and statistics of the data sets.
All data sets are publicly available.\footnote{\label{sociopatterns}\url{https://www.sociopatterns.org}}\footnote{\label{snap}\url{https://snap.stanford.edu}\hfil}\footnote{\label{netwrep}\url{https://networkrepository.com/dynamic.php}}\footnote{\label{konect}\url{http://konect.cc/networks}}
The transition times of the temporal edges are one for all data sets.
\begin{table}
    \centering
    \caption{Statistics of the data sets ($\mathcal{T}(\tg)=\{t\mid (u,v,t,\lambda)\in \tge\}$).} 
    \label{table:datasets_stats2}
    \resizebox{1\linewidth}{!}{ \renewcommand{\arraystretch}{0.9} \setlength{\tabcolsep}{2pt}
        \begin{tabular}{lrrrrrcc}\toprule
            \multirow{4}{0.5cm}{\vspace*{4pt}\textbf{Data~set}\vspace*{4pt}}&\multicolumn{7}{c}{\textbf{Properties}}\\
            \cmidrule{2-8}
            \textbf{ }        &   $|V(\tg)|$         & $|E(\tg)|$  &  $|\mathcal{T}(\tg)|$ &avg.~out-deg.& $\maxdeg^+(\tg)$  & Domain & Ref.
            \\ \midrule
            \emph{Hospital}${}^1$   & 75          & 32\,424      & 9\,453     &   864.6  & 4\,286      & face-to-face          & \cite{vanhems2013estimating}\\
            \emph{Malawi}${}^1$     & 84          & 102\,261     & 43\,436    & 2\,434.8 & 7\,932      & face-to-face          & \cite{ozella2021using}\\
            \emph{Workplace}${}^1$  & 92          & 9\,827       & 7\,104     &    213.6 & 1\,091      & face-to-face          & \cite{genois2018can}\\
            \emph{HTMLConf}${}^1$   & 113         & 20\,818      & 5\,246     &    368.5 & 1\,483      & face-to-face          & \cite{Isella2011}\\
            \emph{Highschool}${}^1$ & 327         & 188\,508     & 7\,375     & 1\,153.0 & 4\,647      & face-to-face          & \cite{mastrandrea2015contact}\\
            \emph{Email}${}^2$      & 1\,866      & 31\,727      & 18\,682    &     17.0 & 2\,635      & communication         & \cite{nr}\\
            \emph{FacebookMsg}${}^3$    & 1\,899      & 59\,798      & 58\,911    &     31.5 & 1\,091      & communication         & \cite{panzarasa2009patterns}\\
            \emph{Infectious}${}^1$ & 10\,972     & 415\,912     & 76\,944    &     75.8 & 616         & face-to-face          & \cite{Isella2011}\\ 
            \emph{FacebookWall}${}^4$   & 63\,731     & 817\,035     & 333\,924   &     25.6 & 1\,098      & messages on user page & \cite{viswanath2009evolution}\\
            \emph{Enron}${}^4$      & 86\,806     & 1\,133\,968  & 213\,167   &     13.1 & 38\,635     & communication         & \cite{klimt2004enron}\\ 
            \emph{AskUbuntu}${}^2$  & 134\,035    & 257\,305     & 257\,079   &      1.9 & 2\,358      & question answering    & \cite{paranjape2017motifs}\\ 
            \emph{Digg}${}^4$       & 279\,630    & 1\,731\,652  & 6\,865     &      6.2 & 995         & friendship            & \cite{hogg2012social}\\ 
            \emph{Wikipedia}${}^4$  & 1\,870\,709 & 39\,953\,145 & 2\,198     &     21.4 & 6\,975      & co-editing            & \cite{mislove2009online}\\  
            \emph{Dblp}${}^4$       & 1\,821\,070 & 11\,859\,561 & 49         &     13.0 & 2\,606      & co-authorship         & \cite{DBLP}\\ 
            \emph{Flickr}${}^3$     & 2\,302\,925 & 33\,140\,016 & 134        &     14.4 & 26\,367     & friendship            & \cite{mislove2008growth}\\
            \emph{Youtube}${}^3$    & 3\,223\,585 & 9\,375\,374  & 203        &      5.8 & 91\,751     & friendship            & \cite{MisloveMGDB07}\\
            \bottomrule
        \end{tabular}
    }
\end{table}

\noindent
\textbf{Algorithms:}
We implemented the following algorithms in C++ using the GNU CC Compiler 10.3.0.
\begin{itemize}
    \item \algr is the implementation of naive recursive algorithm~\Cref{alg:topdown} (see~\Cref{appendix:computation}). 
    \item \algs is the implementation of~\Cref{alg:streaming}. 
\end{itemize}
All experiments were performed on a computer cluster. Each experiment had an exclusive node with an Intel(R) Xeon(R) Gold 6130 CPU @ 2.10GHz and 192 GB of RAM. The source code is available at \url{https://gitlab.com/tgpublic/tgh}.

\subsection{Running Time and Memory Usage}\label{sec:efficiency}
\begin{table*}[htb]
    \centering
    \caption{Running times in seconds (s). OOT: out of time (time limit 12h).} 
    \vspace{-2mm}
    \label{table:runningtime}
    \resizebox{1\linewidth}{!}{ 	\renewcommand{\arraystretch}{0.8}%
        \begin{tabular}{lrrrrrrrrrrrrrrrr}
            \toprule
            \multirow{3}{0.5cm}{\vspace*{4pt}\textbf{Data~set}\vspace*{4pt}}&\multicolumn{2}{c}{$n=1$}&\multicolumn{2}{c}{$n=2$} & \multicolumn{2}{c}{$n=4$}& \multicolumn{2}{c}{$n=8$}& \multicolumn{2}{c}{$n=16$}& \multicolumn{2}{c}{$n=32$}& \multicolumn{2}{c}{$n=64$}& \multicolumn{2}{c}{$n=128$}\\
            \cmidrule(lr){2-3} \cmidrule(lr){4-5} \cmidrule(lr){6-7} \cmidrule(lr){8-9} \cmidrule(lr){10-11}  \cmidrule(lr){12-13} \cmidrule(lr){14-15} \cmidrule(lr){16-17}  
            \textbf{ }     & \algr & \algs & \algr & \algs & \algr & \algs & \algr & \algs & \algr & \algs & \algr & \algs & \algr & \algs & \algr & \algs \\     
            \midrule
            \emph{Hospital}     & $0.17 $  & $0.00 $ & $5.42 $ & $0.09 $ & $15.97 $ & $0.25 $ & $37.06 $ & $0.57 $ & $78.40 $ & $1.25 $ & $162.55 $ & $2.68 $ & $328.65 $ & $5.92 $ & $659.53 $ & $14.79 $ \\
            \emph{Malawi}       & $1.08 $  & $0.00 $ & $37.59 $ & $0.44 $ & $110.09 $ & $1.37 $ & $255.22 $ & $3.43 $ & $545.44 $ & $7.96 $ & $1125.34 $ & $18.83 $ & $2272.54 $ & $42.75 $ & $4544.73 $ & $107.89 $ \\
            \emph{Workplace}    & $0.02 $  & $0.00 $ & $0.31 $ & $0.01 $ & $0.88 $ & $0.02 $ & $2.03 $ & $0.04 $ & $4.31 $ & $0.10 $ & $8.83 $ & $0.20 $ & $17.97 $ & $0.46 $ & $37.45 $ & $0.94 $ \\
            \emph{HTMLConf}     & $0.06 $  & $0.00 $ & $1.12 $ & $0.02 $ & $3.26 $ & $0.06 $ & $7.56 $ & $0.14 $ & $16.17 $ & $0.31 $ & $33.43 $ & $0.67 $ & $67.60 $ & $1.45 $ & $135.38 $ & $3.25 $ \\
            \emph{Highschool}   & $0.97 $  & $0.01 $ & $28.84 $ & $0.52 $ & $85.15 $ & $1.46 $ & $197.26 $ & $3.51 $ & $420.99 $ & $8.49 $ & $864.52 $ & $20.58 $ & $1747.02 $ & $46.36 $ & $3506.29 $ & $100.60 $ \\
            \emph{Email}        & $0.04 $  & $0.00 $ & $0.74 $ & $0.02 $ & $2.10 $ & $0.04 $ & $4.79 $ & $0.09 $ & $10.27 $ & $0.19 $ & $21.74 $ & $0.39 $ & $45.99 $ & $0.84 $ & $109.81 $ & $1.86 $ \\
            \emph{FacebookMsg}  & $0.05 $  & $0.00 $ & $0.43 $ & $0.02 $ & $1.14 $ & $0.04 $ & $2.56 $ & $0.08 $ & $5.52 $ & $0.15 $ & $11.94 $ & $0.31 $ & $26.70 $ & $0.64 $ & $63.89 $ & $1.48 $ \\
            \emph{Infectious}   & $0.53 $  & $0.05 $ & $3.07 $ & $0.19 $ & $8.35 $ & $0.46 $ & $18.81 $ & $1.02 $ & $39.87 $ & $1.98 $ & $76.73 $ & $4.19 $ & $144.48 $ & $8.51 $ & $270.94 $ & $17.48 $ \\
            \emph{FacebookWall} & $0.92 $  & $0.19 $ & $4.99 $ & $0.67 $ & $13.18 $ & $1.52 $ & $31.11 $ & $3.48 $ & $69.01 $ & $6.03 $ & $135.03 $ & $11.31 $ & $310.44 $ & $22.49 $ & $575.94 $ & $43.46 $ \\
            \emph{Enron}        & $2.35 $  & $0.08 $ & $62.63 $ & $0.81 $ & $184.38 $ & $2.07 $ & $411.59 $ & $4.89 $ & $866.53 $ & $11.03 $ & $1882.11 $ & $24.45 $ & $4226.25 $ & $52.33 $ & $9579.23 $ & $122.79 $ \\
            \emph{AskUbuntu}    & $0.10 $  & $0.04 $ & $0.35 $ & $0.07 $ & $0.69 $ & $0.12 $ & $1.23 $ & $0.21 $ & $2.50 $ & $0.38 $ & $5.31 $ & $0.72 $ & $13.15 $ & $1.44 $ & $20.43 $ & $3.09 $ \\
            \emph{Digg}         & $1.19 $  & $0.22 $ & $9.71 $ & $0.71 $ & $26.49 $ & $1.61 $ & $62.80 $ & $3.33 $ & $120.30 $ & $6.84 $ & $229.93 $ & $13.83 $ & $364.88 $ & $27.61 $ & $467.20 $ & $54.81 $ \\
            \emph{Wikipedia}    & $67.36 $ & $7.86 $ & $855.44 $ & $26.85 $ & $2207.03 $ & $61.29 $ & $4863.44 $ & $117.03 $ & $10332.81 $ & $230.70 $ & $21998.44 $ & $457.65 $ & OOT & $861.95 $ & OOT & $1794.60 $ \\
            \emph{Dblp}         & $10.78 $ & $6.64 $ & $34.39 $ & $19.58 $ & $78.12 $ & $35.92 $ & $168.25 $ & $68.18 $ & $349.69 $ & $124.74 $ & $708.33 $ & $227.67 $ & $867.77 $ & $419.43 $ & $887.72 $ & $827.37 $ \\
            \emph{Flickr}       & $19.90 $ & $7.27 $ & $154.39 $ & $32.61 $ & $397.31 $ & $82.62 $ & $870.88 $ & $168.92 $ & $1767.10 $ & $332.29 $ & $3323.15 $ & $640.81 $ & $5373.19 $ & $1282.84 $ & $6310.73 $ & $2626.10 $ \\
            \emph{Youtube}      & $14.09 $ & $5.38 $ & $82.18 $ & $48.30 $ & $204.97 $ & $135.38 $ & $454.32 $ & $323.65 $ & $973.77 $ & $739.84 $ & $2108.57 $ & $1657.65 $ & $4519.12 $ & $3548.93 $ & $7340.79 $ & $7316.00 $ \\
            
            \bottomrule
        \end{tabular}
    }
    \vspace{-2mm}
\end{table*}
\begin{figure}\centering\centering\setlength{\belowcaptionskip}{-4pt}
    \begin{subfigure}{0.5\linewidth}
        \includegraphics[width=\linewidth]{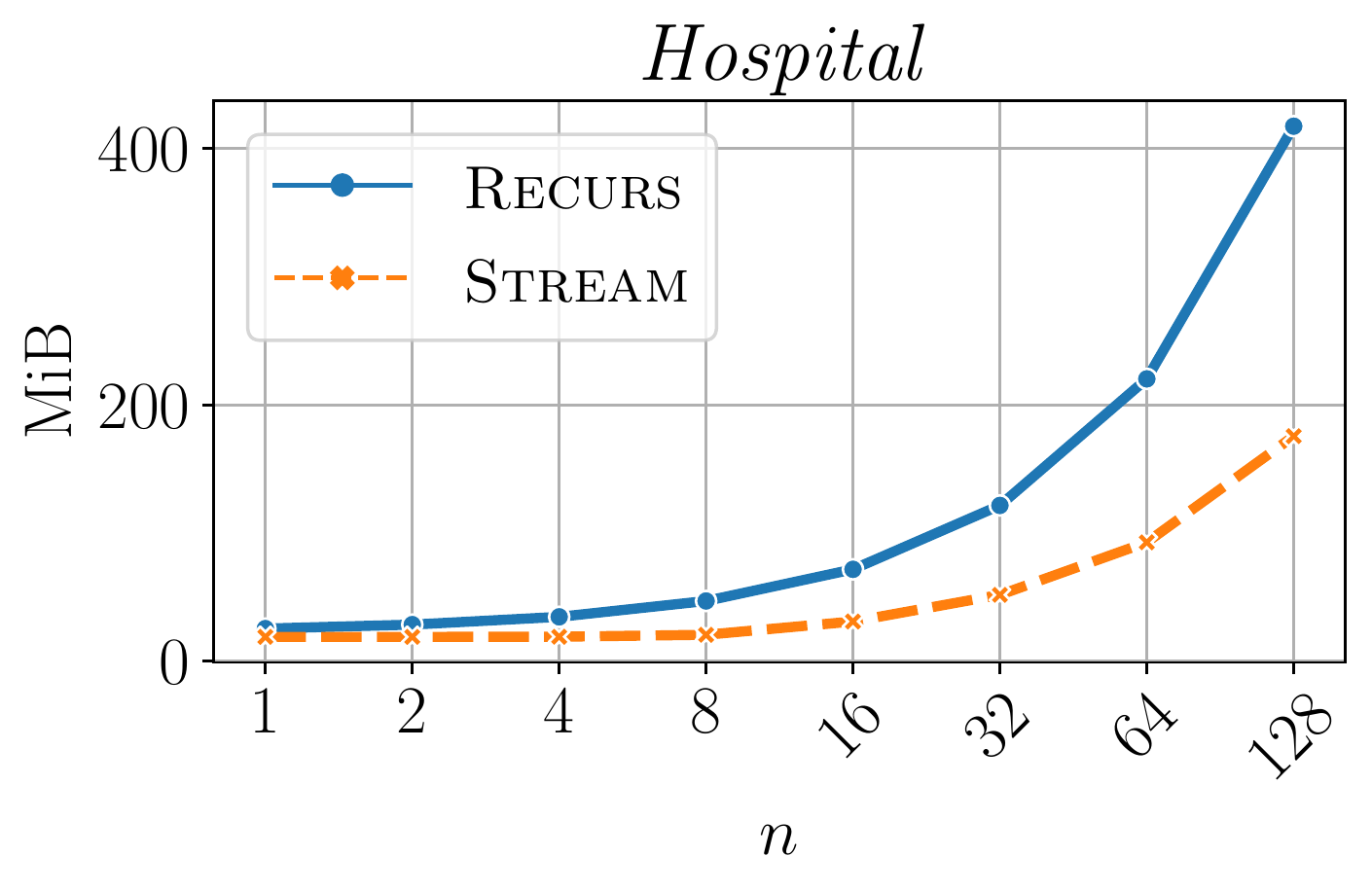}
        \label{fig:memhospital}
    \end{subfigure}\hfil%
    \begin{subfigure}{0.5\linewidth}
        \includegraphics[width=\linewidth]{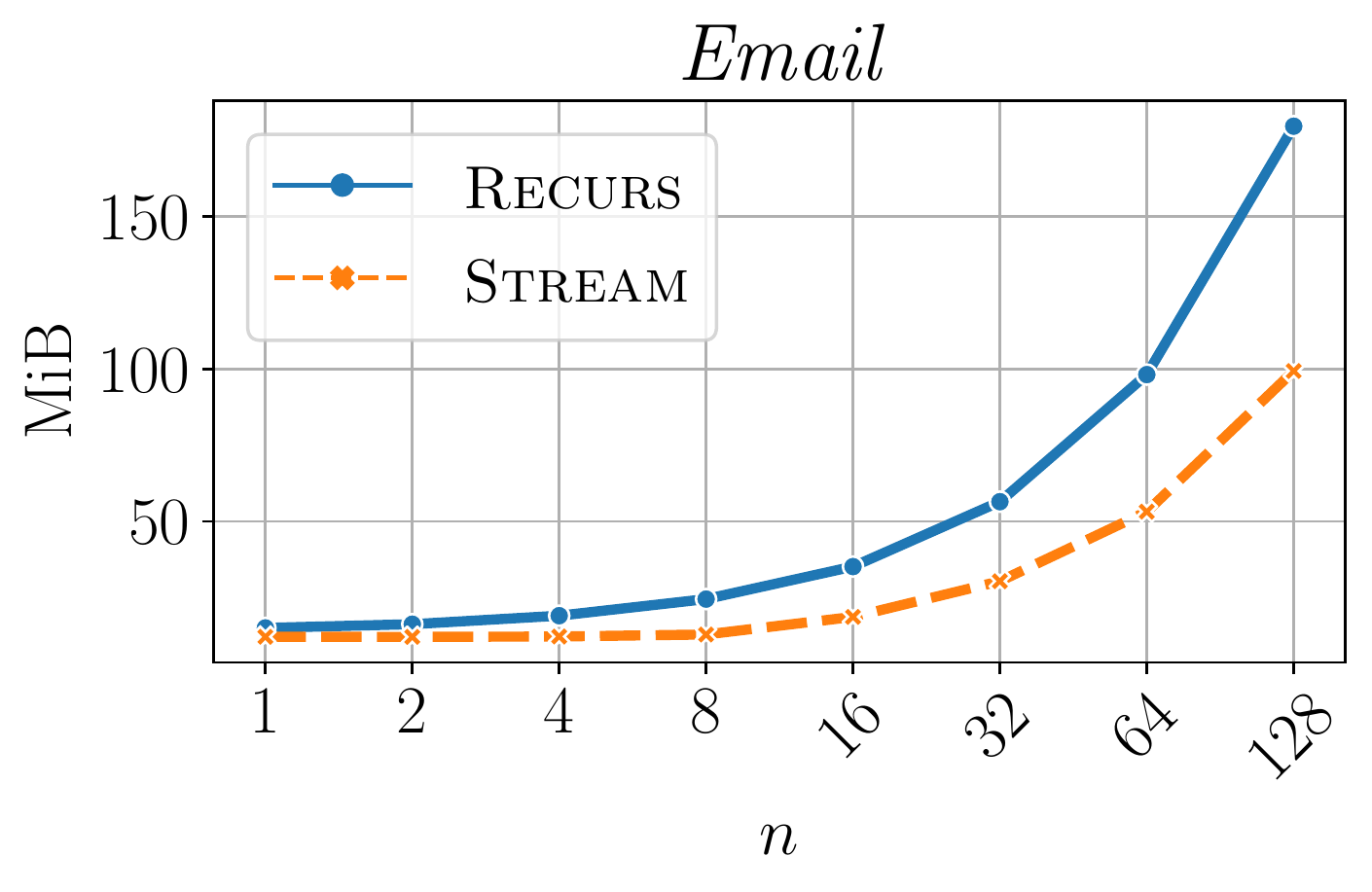}
        \label{fig:mememail}
    \end{subfigure}
    \vspace{-8mm}
    \caption{Effect of increasing $n$ on the memory.} %
    \label{fig:mem}
\end{figure}
\begin{figure}\centering\setlength{\belowcaptionskip}{-4mm}
    \begin{subfigure}{0.5\linewidth}
        \includegraphics[width=\linewidth]{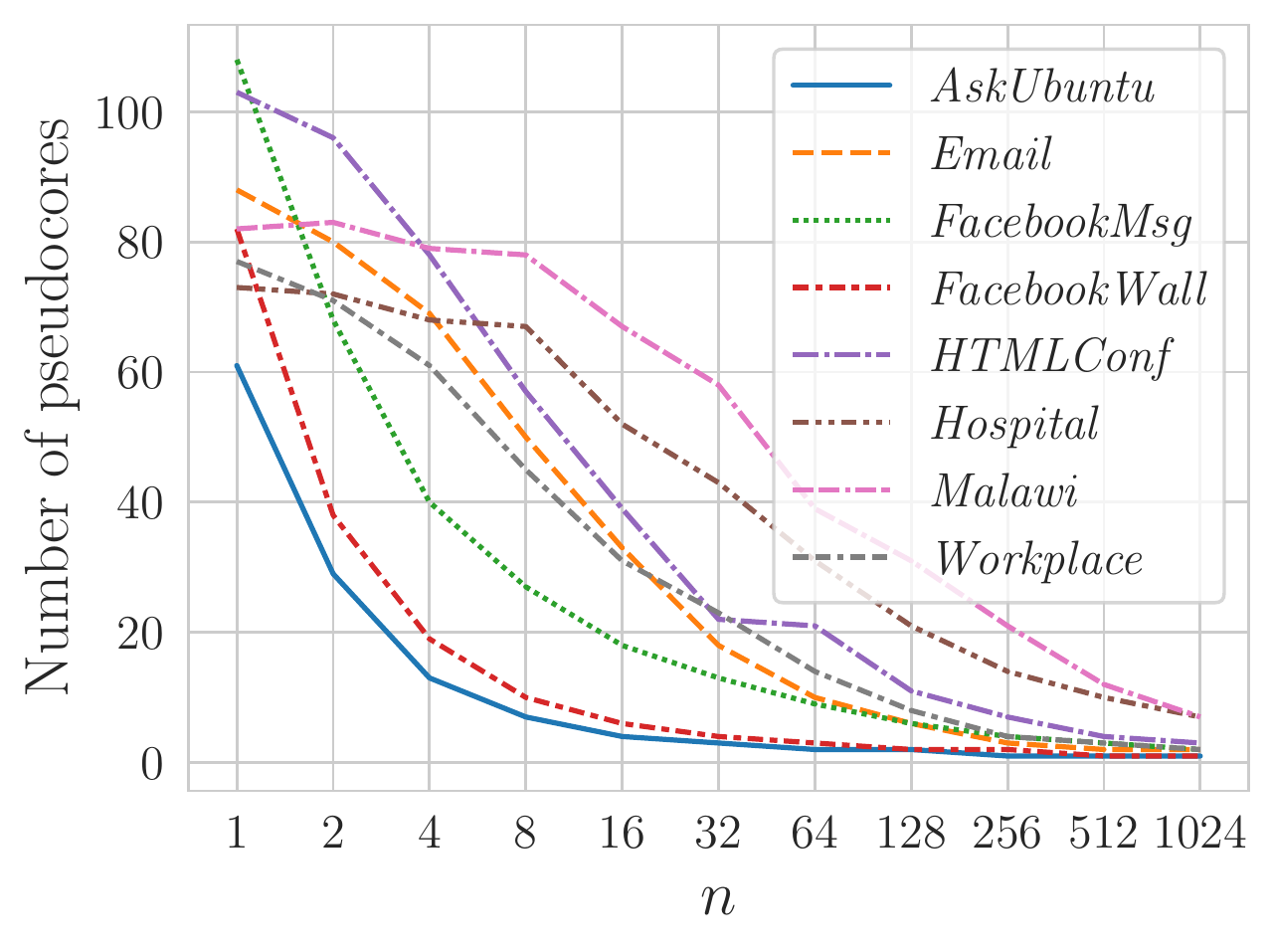}
        \caption{}
        \label{fig:coresa}
    \end{subfigure}\hfil%
    \begin{subfigure}{0.5\linewidth}
        \includegraphics[width=\linewidth]{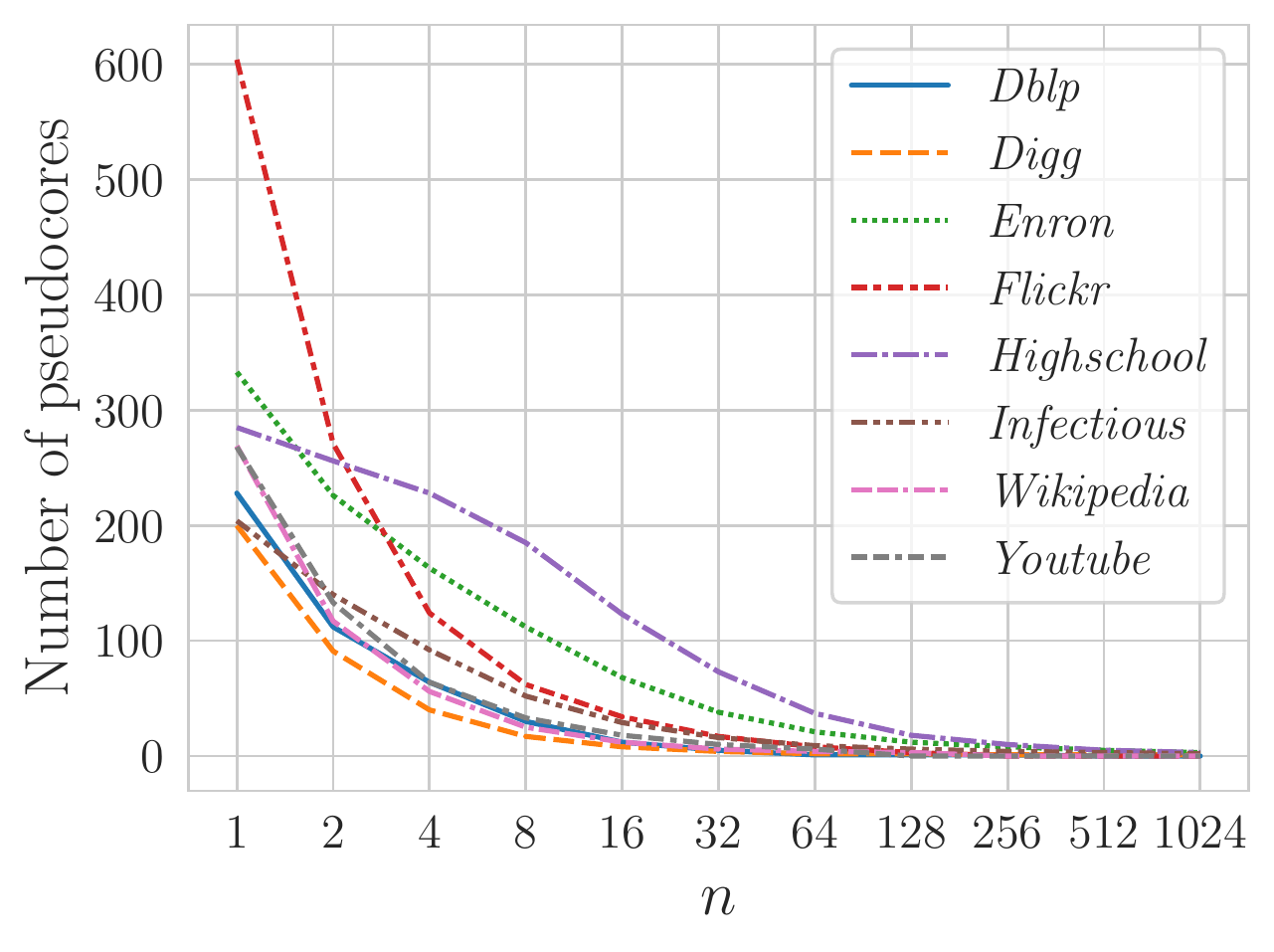}
        \caption{}
        \label{fig:coresb}
    \end{subfigure}\hfil%
    \caption{Effect of increasing $n$ on the number of pseudocores. %
    }
    \label{fig:cores}
\end{figure}
\begin{figure*}\centering\setlength{\belowcaptionskip}{-2pt}
    \begin{subfigure}{0.2\linewidth}\hfill%
        \includegraphics[width=\linewidth]{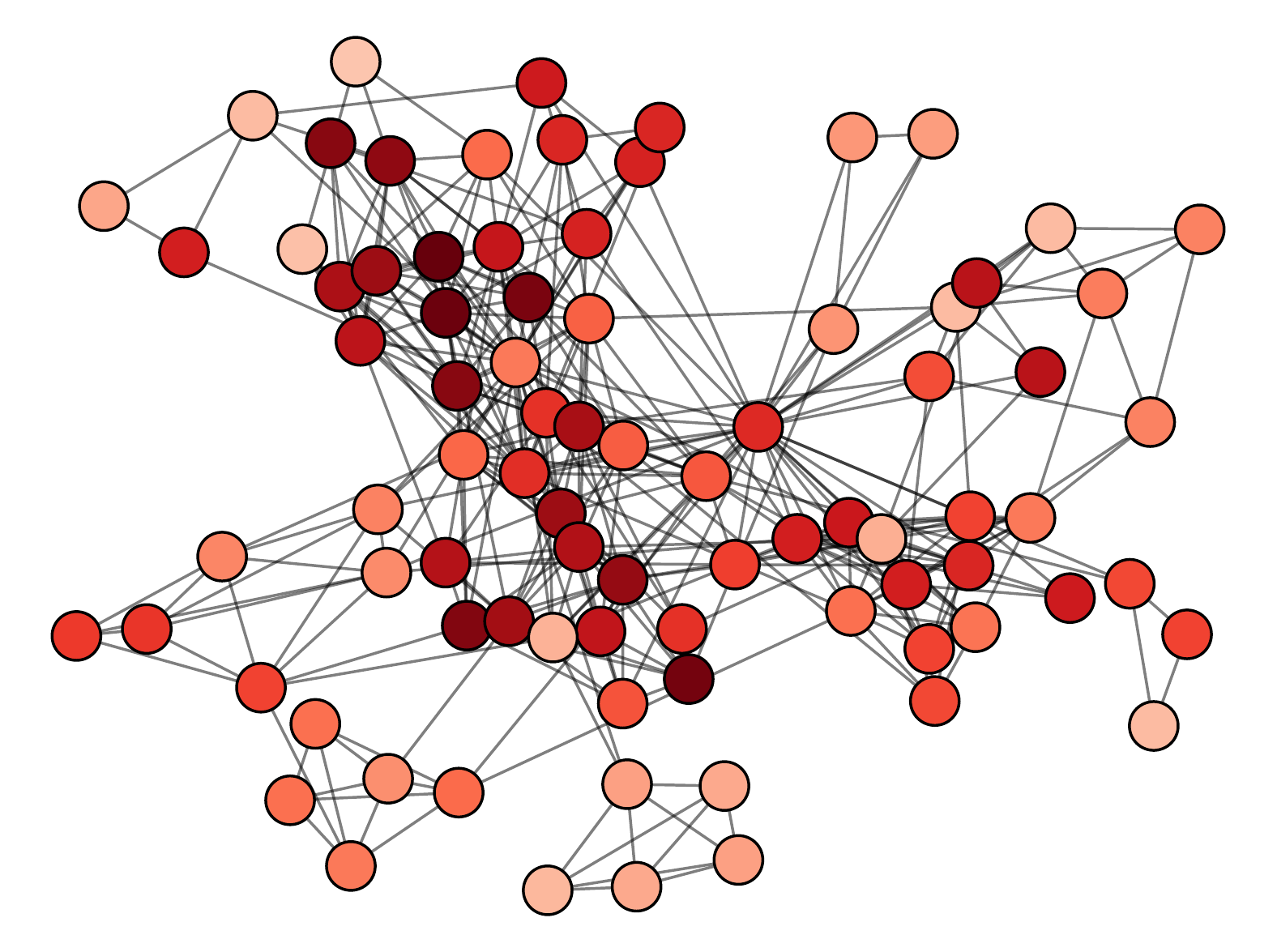}
        \caption{$n=32$-th Temporal H-index with $58$ different ranks}
        \label{fig:pseudocorea}
    \end{subfigure}\hfill%
    \begin{subfigure}{0.2\linewidth}\hfill%
        \includegraphics[width=\linewidth]{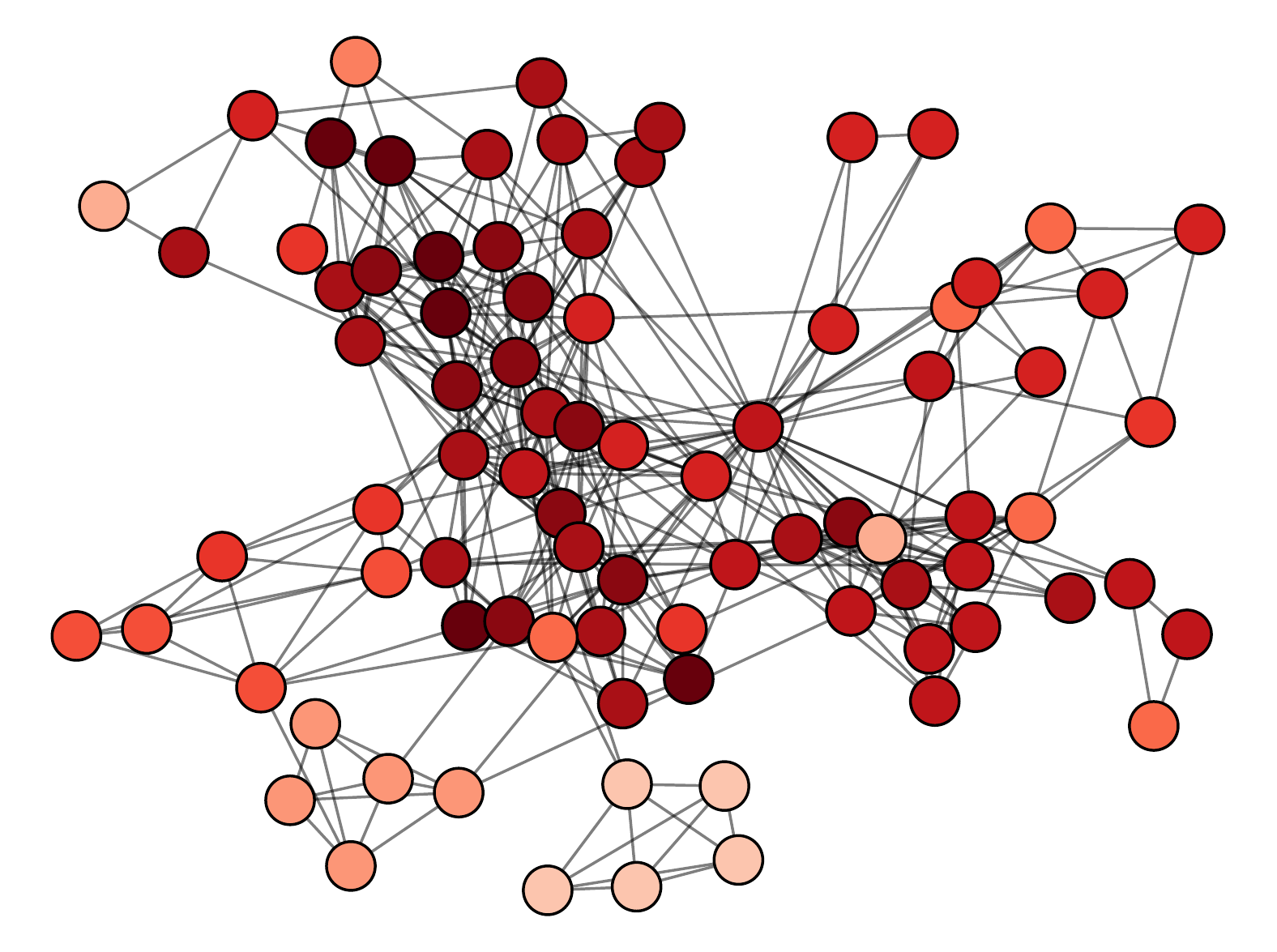}
        \caption{$n=512$-th Temporal H-index with $12$ different ranks}
        \label{fig:pseudocoreb}
    \end{subfigure}\hfill%
    \begin{subfigure}{0.2\linewidth}
        \includegraphics[width=\linewidth]{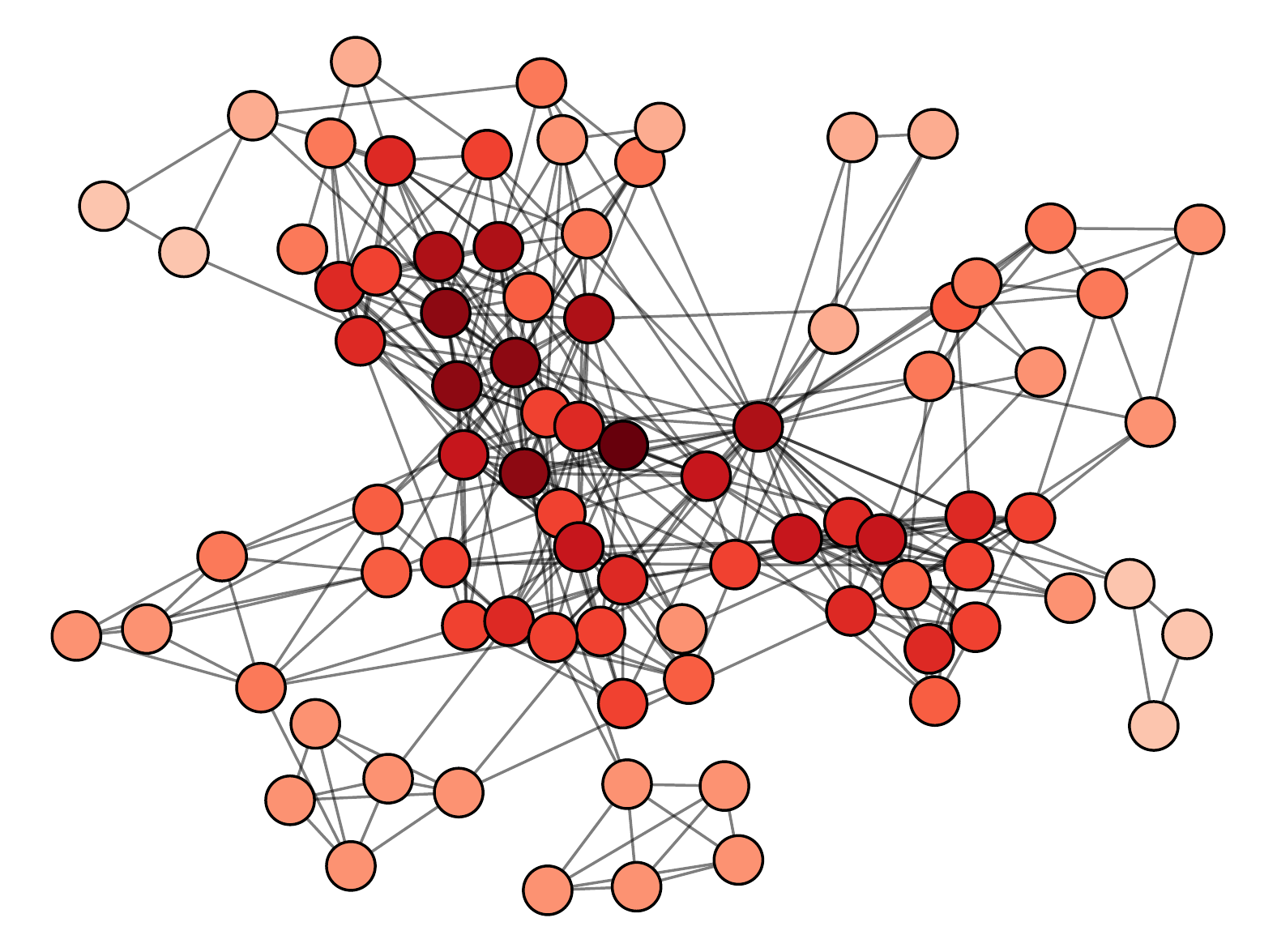}
        \caption{Static H-index with $11$ different ranks}
        \label{fig:pseudocorec}
    \end{subfigure}\hfill%
    \begin{subfigure}{0.2\linewidth}
        \includegraphics[width=\linewidth]{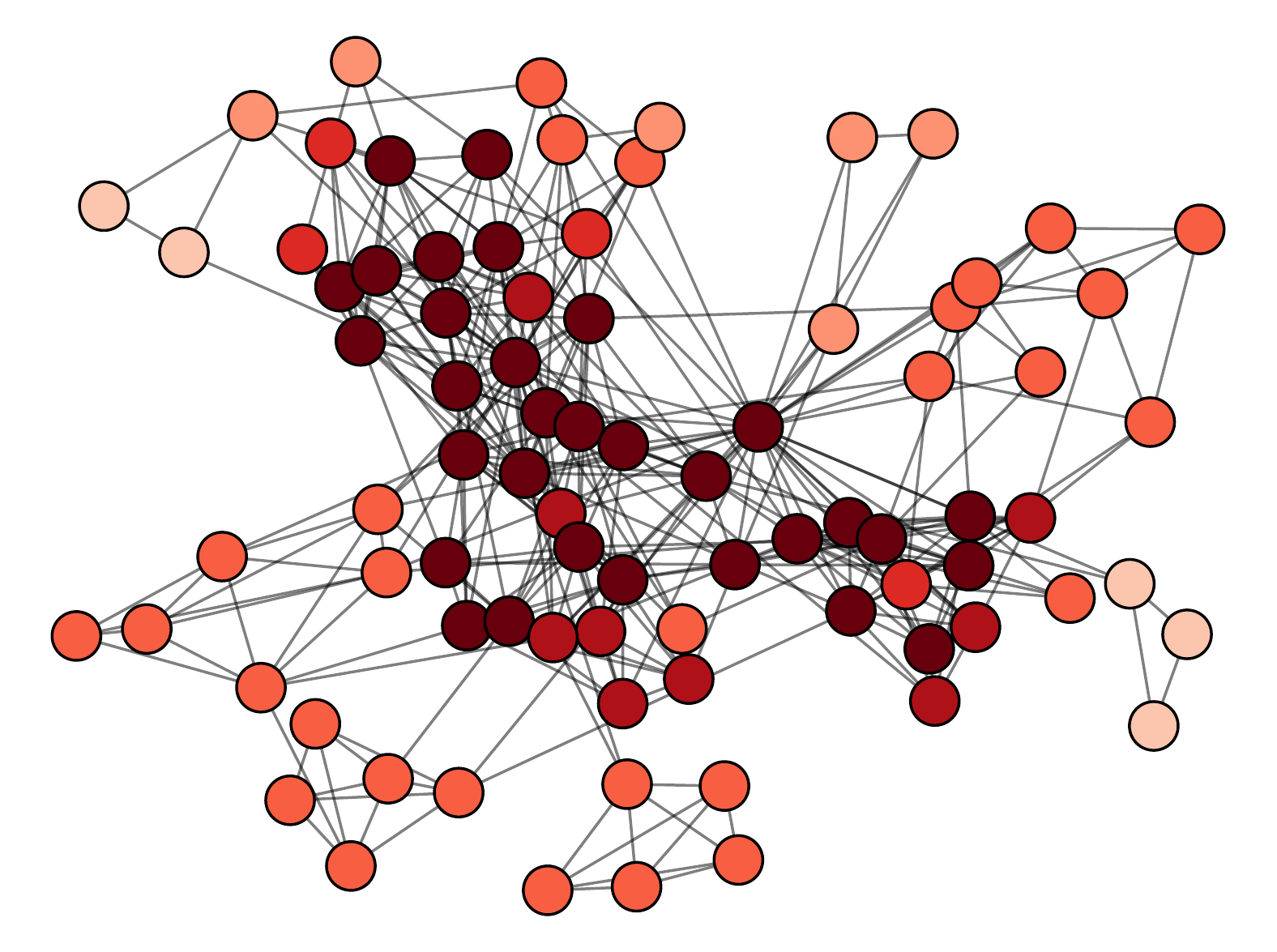}
        \caption{Static $k$-core with seven different ranks}
        \label{fig:pseudocored}
    \end{subfigure}
    \caption{Comparison of the outward $n$-th temporal H-index, the static H-index, and the static $k$-core values where a darker color means a higher ranking. The shown network is based on human face-to-face contacts in a village in \emph{Malawi}. }
    \label{fig:pseudocore}
\end{figure*}
\begin{figure*}\centering\setlength{\belowcaptionskip}{-6pt}
    \hspace{4mm}\includegraphics[width=0.475\linewidth]{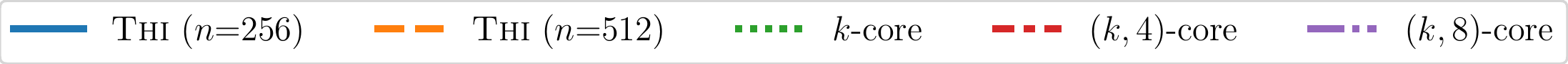}\hfill%
    \includegraphics[width=0.475\linewidth]{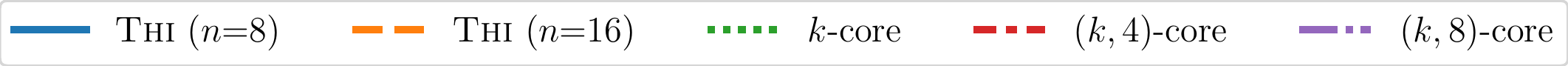}
    
    \begin{subfigure}{0.25\linewidth}\hfill%
        \includegraphics[width=\linewidth]{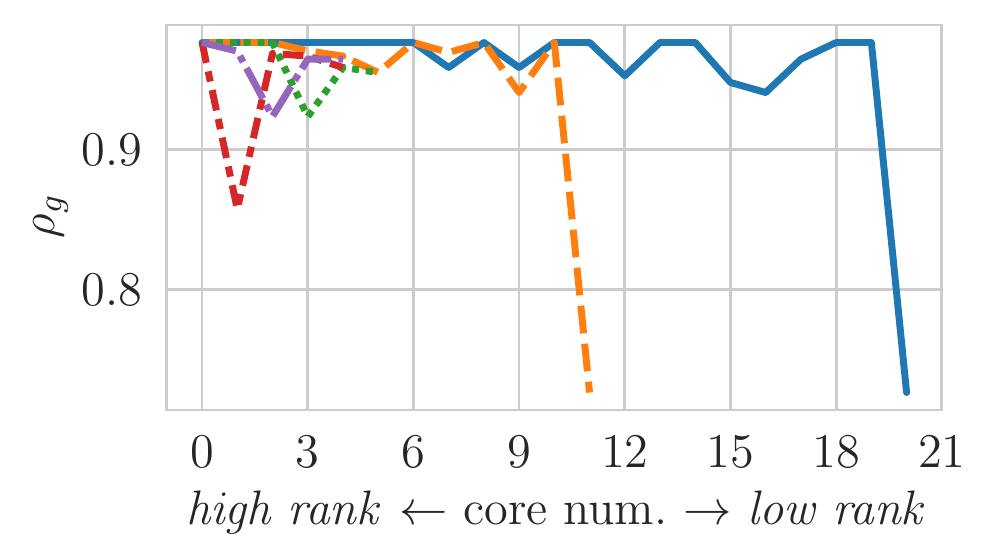}
        \caption{\emph{Malawi -- global score}}
        \label{fig:reacha}
    \end{subfigure}\hfill%
    \begin{subfigure}{0.25\linewidth}\hfill%
        \includegraphics[width=\linewidth]{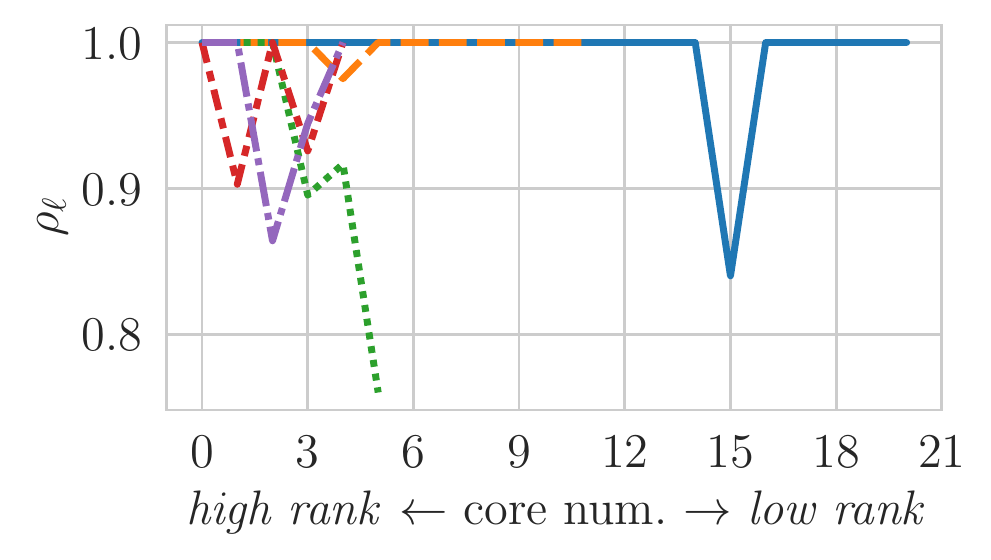}
        \caption{\emph{Malawi -- local score}}
        \label{fig:reachb}
    \end{subfigure}\hfill%
    \begin{subfigure}{0.25\linewidth}
        \includegraphics[width=\linewidth]{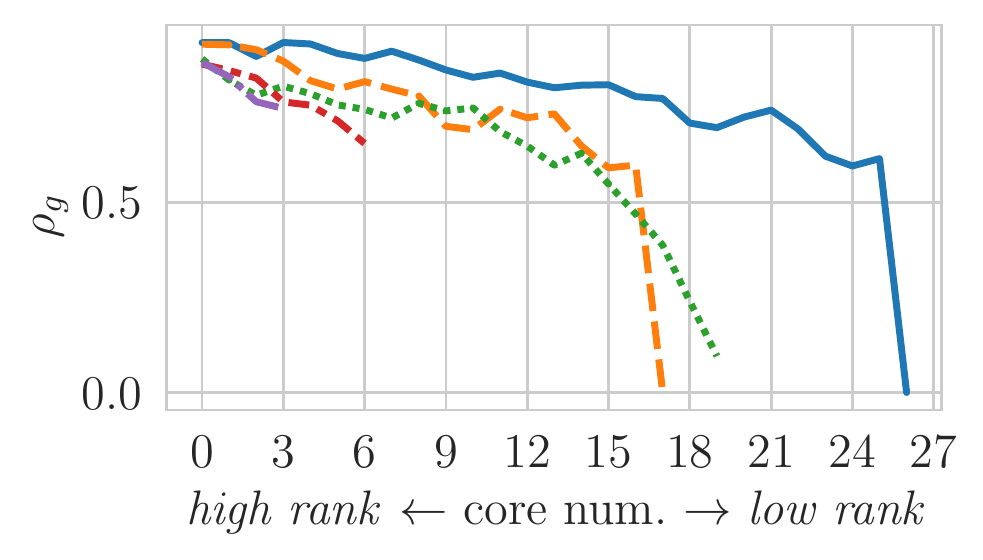}
        \caption{\emph{FacebookMsg -- global score}}
        \label{fig:reachc}
    \end{subfigure}\hfill%
    \begin{subfigure}{0.25\linewidth}
        \includegraphics[width=\linewidth]{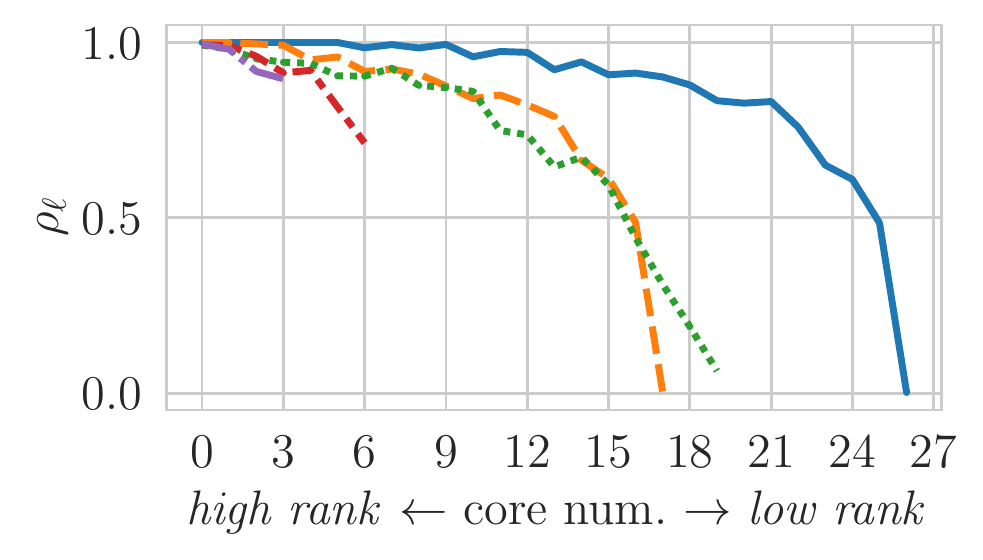}
        \caption{\emph{FacebookMsg -- local score}}
        \label{fig:reachd}
    \end{subfigure}
    \caption{Comparison of global and local reachability scores of different network decompositions.}
    \label{fig:reachability}
\end{figure*}

\Cref{table:runningtime} shows the running times for \algr and \algs for $n\in\{2^i\mid i\in\{0,\ldots 7\} \}$.
We used a time limit of twelve hours.
Our streaming algorithm \algs performs very well even on data sets with tens of millions of edges, e.g., the \emph{Wikipedia} data set.
\algs is faster than \algr for all datasets and all $n$. In most cases, the speed-up is significant and up to several orders of magnitude.
In general, the running time increases roughly linearly with increasing $n$, as expected for both algorithms.
Recall that \algs computes the $j$-th order H-index for all $0\leq j\leq n$, where \algr only computes the value for $j=n$. Hence,  computing the remaining values for $0\leq j\leq n-1$ for \algr would require additional running time.
As expected, the speed-up of \algs compared to \algr is higher for the data sets with a high average degree because the asymptotic running time of \algr contains the node degree as a quadratic factor (see \Cref{theorem:algrec}), whereas for \algs it has the node degree as a linear factor (see~\Cref{theorem:algstream}).
\algr cannot finish the computations for the \emph{Wikipedia} data set and $n\geq 64$ within the time limit.
\algs has competitive running times compared to other state-of-the-art temporal centrality measures and core-decompositions---a further discussion can be found in \Cref{appendix:runningtimes}.
The memory usage of both algorithms asymptotically behaves similarly.
\Cref{fig:mem} compares the memory usage for \emph{Hospital} and \emph{Email}. 
\algr needs slightly more memory because the multisets $L$ are maintained on the stack for recursive calls.

\subsection{Additional Efficiency Comparisons}\label{appendix:runningtimes}
We additionally compare the running times of our new approach to related centrality measures and core decompositions.
We provide the running times of the following methods:
\begin{itemize}
    \item \textsc{H-Index} is the static H-index and \textsc{$k$-core} is the static $k$-core. Both were implemented in C++.
    \item \textsc{Tc}, \textsc{Twc}, \textsc{Tkatz}, and \textsc{Tpr} are the temporal closeness, walk, Katz, and PageRank centrality. We used the C++ implementations provided by~\cite{oettershagen2022tglib}.
    \item \textsc{Tb} is the temporal betweenness centrality using strict temporal path (i.e., temporal paths that are strictly advancing in time for each edge) from~\cite{BussMNR20}. The C++ implementation was provided by the authors.
    \item \textsc{$(k,h)$-core} with $h\in\{2,4,8\}$ is the $(k,h)$-core described in \cite{wu2015core}. We used the C++ implementations provided by~\cite{oettershagen2022tglib}.
    \item \textsc{Span-core} is temporal core decomposition from~\cite{galimberti2020span}. We used the Cython implementation computing the maximum span-cores provided by the authors.  
\end{itemize}
See \Cref{tab:overview2} in the appendix for an overview of the time and space complexities of the centrality measures and core decompositions showing that our $n$-temporal H-index has competitive complexities. Which is also verified by the following empirical evaluation.
\Cref{table:runningtime2} shows the running times in seconds.
We see that our streaming algorithm for the temporal H-index is for low $n$, often significantly faster than the algorithms for the computation of the more demanding temporal closeness and betweenness centrality measures. 
For $n=1$, \algs has similar running times as computing the static H-index in the underlying aggregated graph, where the temporal variant is slightly faster in almost all cases. Our algorithm needs more time than the temporal Katz and PageRank computations, which is expected as the former are computable in linear time (see \Cref{tab:overview2} in the appendix). 
In the case of the core decompositions, our streaming algorithm \algs has higher running times compared to the static $k$ core and the $(k,h)$-cores. The reason is that the latter can be computed in linear time. The implementation of the span cores cannot finish the computations for most data sets due to out-of-memory errors. The reason is that the algorithm needs memory in length of the time interval spanned by the temporal graph.
\begin{table*}[htb]
    \centering
    \caption{Running times of related centrality measures and core decompositions in seconds (s). OOT: Out of time (time limit 12h). OOM: Out of memory (available memory 196GB).} 
    \label{table:runningtime2}
    \resizebox{1\linewidth}{!}{ 	\renewcommand{\arraystretch}{0.8}\setlength{\tabcolsep}{2mm}
        \begin{tabular}{lrrrrrrrrrrrr}
            \toprule
            \multirow{3}{0.5cm}{\vspace*{2pt}\textbf{Data~set}\vspace*{2pt}}&\multicolumn{6}{c}{\textbf{Centrality Measures}}&&\multicolumn{5}{c}{\textbf{Core Decompositions}} \\
            \cmidrule(lr){2-8} \cmidrule(lr){9-13}  
            \textbf{ }           & \textsc{H-Index} & \textsc{Tc} & \textsc{Tb} &\textsc{Twc} & \textsc{Tkatz} & \textsc{Tpr} &  & \textsc{$k$-core}  & \textsc{$(k,2)$-core} & \textsc{$(k,4)$-core} & \textsc{$(k,8)$-core} & \textsc{Span-core}  \\     
            \midrule
            \emph{Hospital}     &  0.00 &   0.03 &   90.96 &  1.89 & 0.00 & 0.00 & &  0.00 &   0.00 &  0.00 &  0.00 & 0.67   \\     
            \emph{Malawi}       &  0.00 &   0.04 &  904.13 & 21.85 & 0.00 & 0.00 & &  0.01 &   0.01 &  0.01 &  0.01 & 2.42   \\     
            \emph{Workplace}    &  0.00 &   0.03 &    8.28 &  0.15 & 0.00 & 0.00 & &  0.00 &   0.00 &  0.00 &  0.00 & 1.38   \\     
            \emph{HTMLConf}     &  0.00 &   0.05 &   34.32 &  0.42 & 0.00 & 0.00 & &  0.00 &   0.00 &  0.00 &  0.00 & 0.37   \\     
            \emph{Highschool}   &  0.00 &   0.27 & 2349.55 & 19.52 & 0.00 & 0.01 & &  0.03 &   0.03 &  0.03 &  0.03 & 1.30   \\     
            \emph{Email}        &  0.00 &   0.08 &  109.06 &  0.23 & 0.00 & 0.00 & &  0.01 &   0.01 &  0.01 &  0.01 &  OOM   \\     
            \emph{FacebookMsg}  &  0.00 &   0.23 &  212.94 &  0.20 & 0.00 & 0.00 & &  0.02 &   0.02 &  0.02 &  0.02 &  OOM   \\     
            \emph{Infectious}   &  0.01 &   0.90 &  988.09 &  1.23 & 0.00 & 0.01 & &  0.07 &   0.07 &  0.07 &  0.07 &  OOM   \\     
            \emph{FacebookWall} &  0.48 & 257.02 &     OOM &  3.42 & 0.01 & 0.03 & &  1.62 &   1.08 &  1.08 &  1.08 &  OOM   \\     
            \emph{Enron}        &  0.05 &  63.20 &     OOM & 10.78 & 0.01 & 0.02 & &  0.69 &   0.59 &  0.56 &  0.54 &  OOM   \\     
            \emph{AskUbuntu}    &  0.05 &  32.56 &     OOM &  0.23 & 0.01 & 0.00 & &  0.37 &   0.20 &  0.20 &  0.20 &  OOM   \\     
            \emph{Digg}         &  0.51 & 608.37 &     OOM &  1.74 & 0.00 & 0.04 & &  3.86 &   2.45 &  2.17 &  2.17 &  OOM   \\     
            \emph{Wikipedia}    & 18.46 &    OOT &     OOM &152.32 & 0.47 & 1.17 & &182.75 & 110.48 & 94.44 & 94.33 &  OOM   \\     
            \emph{Dblp}         &  7.82 &    OOT &     OOM & 19.16 & 0.18 & 0.39 & & 24.86 &  14.07 & 11.88 & 11.52 &  OOM   \\     
            \emph{Flickr}       & 11.77 &    OOT &     OOM & 60.24 & 0.28 & 0.85 & & 80.93 &  41.57 & 33.35 & 34.34 &  OOM   \\     
            \emph{Youtube}      &  7.83 &    OOT &     OOM & 26.13 & 0.29 & 0.59 & & 31.39 &  12.97 & 13.02 & 13.14 &  OOM   \\     
            
            \bottomrule
        \end{tabular}
    }
\end{table*}

\begin{figure*}\centering\setlength{\belowcaptionskip}{-6pt}
    \hspace{3mm}\includegraphics[width=0.9\linewidth]{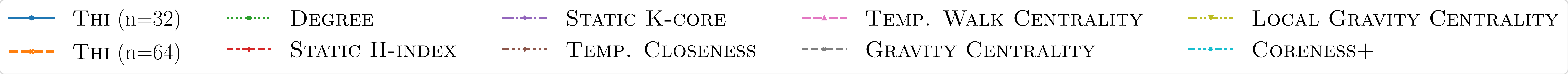}
    \captionsetup[subfigure]{aboveskip=-2pt,belowskip=-2pt}
    \captionsetup[figure]{aboveskip=-2pt,belowskip=-2pt}
    
    \begin{subfigure}{0.24\linewidth}
        \includegraphics[width=\linewidth]{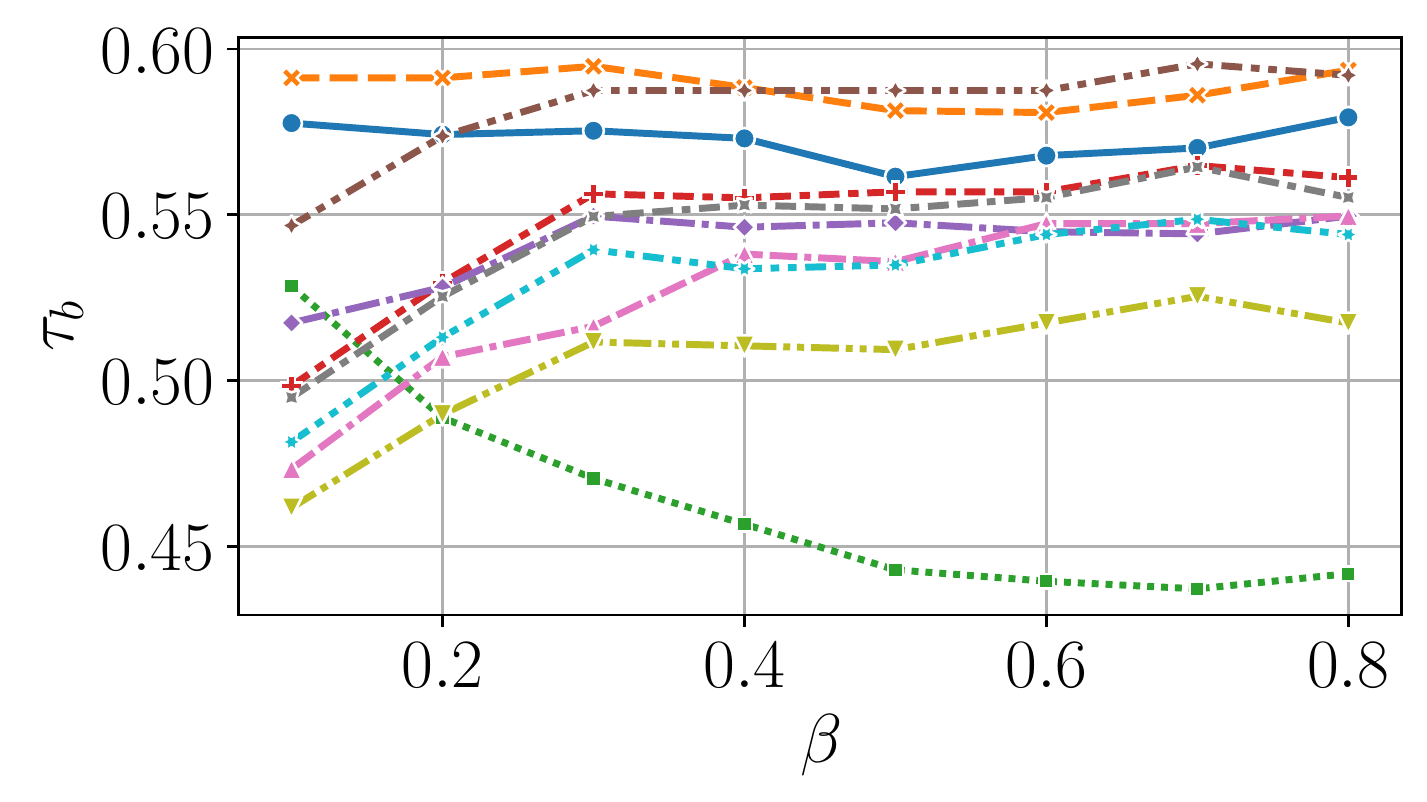}
        \caption{\emph{Malawi}}
    \end{subfigure}\hfil%
    \begin{subfigure}{0.24\linewidth}
        \includegraphics[width=\linewidth]{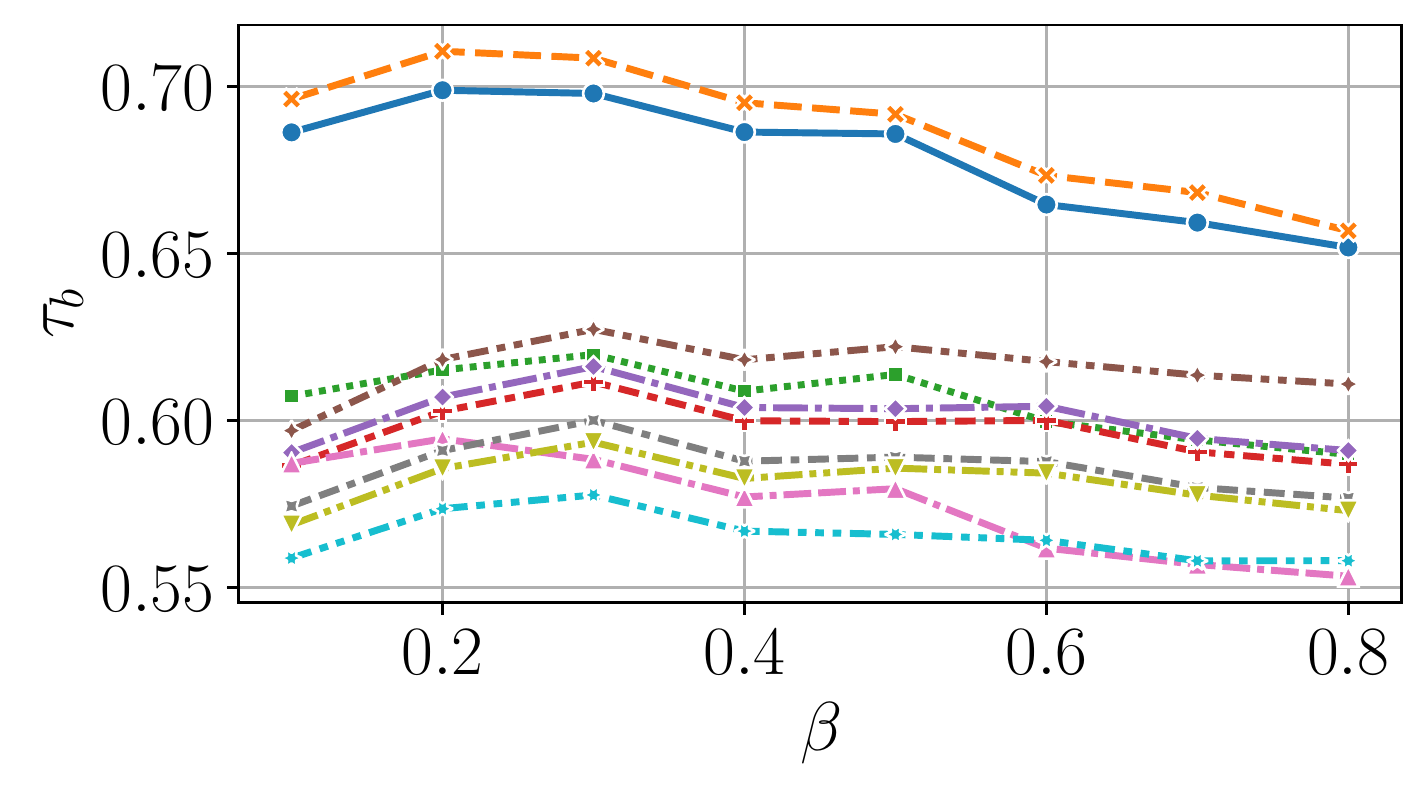}
        \caption{\emph{FacebookMsg}}
    \end{subfigure}\hfil%
    \begin{subfigure}{0.24\linewidth}
        \includegraphics[width=\linewidth]{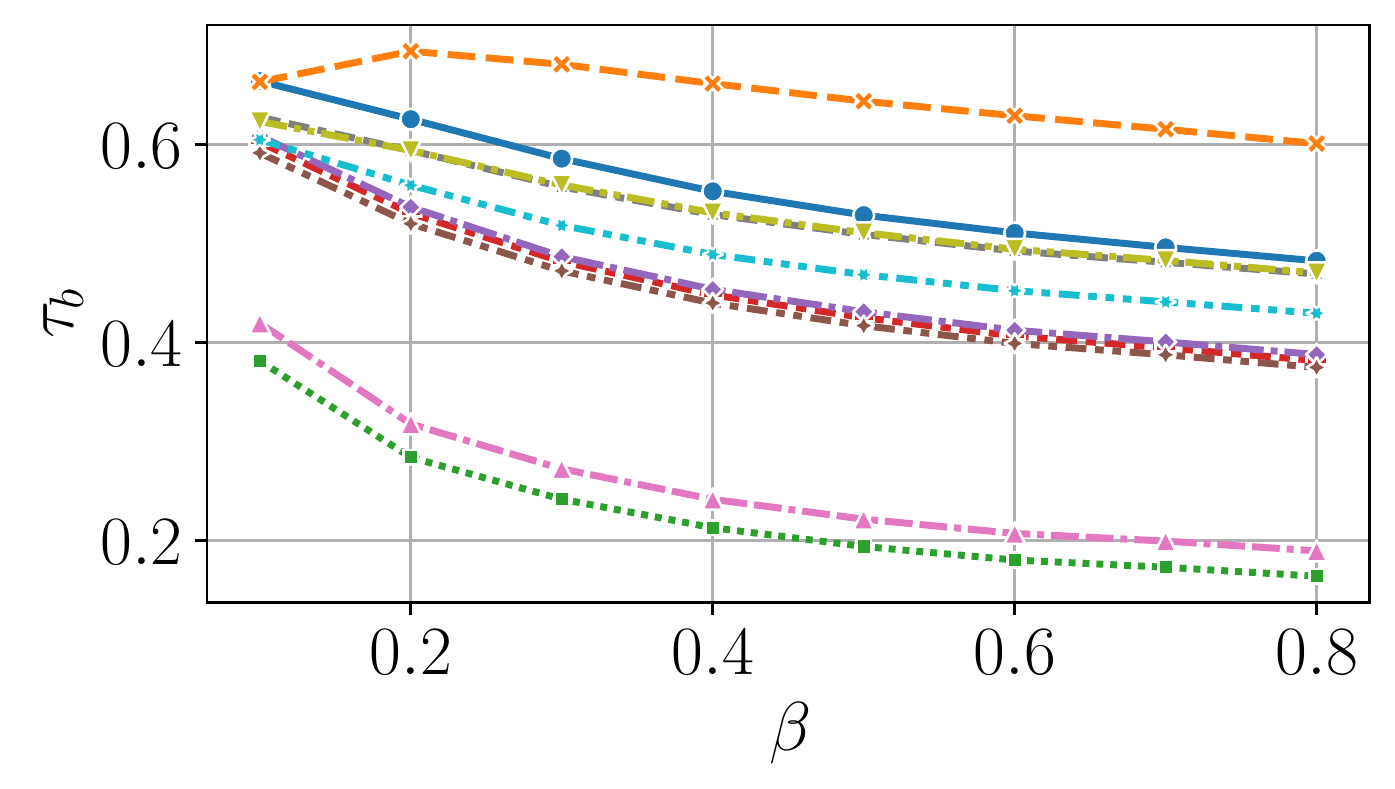}
        \caption{\emph{Infectious}}
    \end{subfigure}\hfil%
    \begin{subfigure}{0.24\linewidth}
        \includegraphics[width=\linewidth]{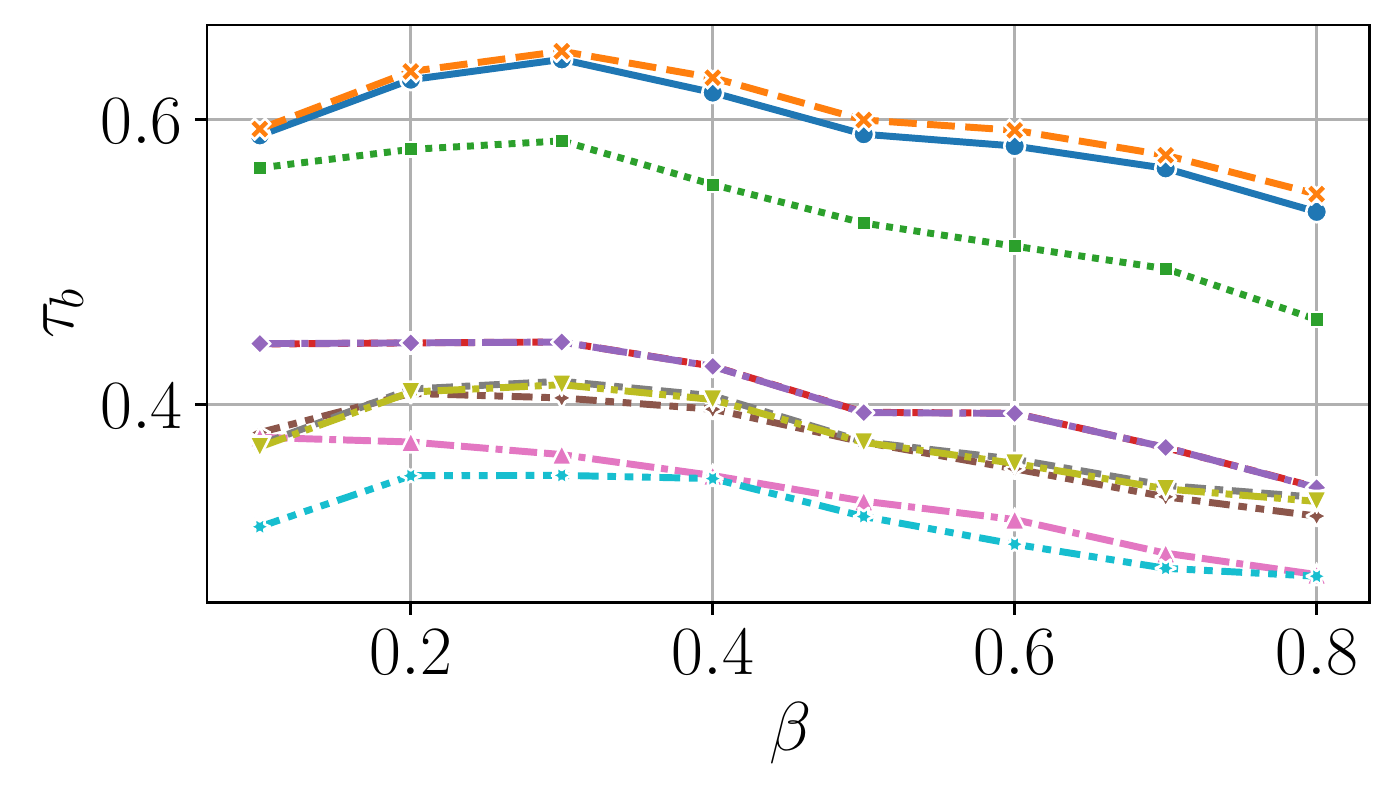}
        \caption{\emph{Email}}
    \end{subfigure}
    \caption{Kendall $\tau_b$ correlations between the rankings of different centrality measures and the ranking according to the SIR model for infection probabilities $\beta\in\{0.1,0.2,\ldots,0.8\}$. \textsc{Thi} denotes the $n$-th order temporal H-index.}
    \label{fig:sir}
\end{figure*}
\subsection{Effect of the Parameter $n$}
\Cref{fig:cores} shows the effect of increasing $n\in\{2^i\mid i\in\{0,\ldots 10\} \}$ on the number of out-pseudocores, i.e., distinct node values for the outward $n$-th order temporal H-index. \Cref{fig:coresa} shows the results for the data sets with less than 150 pseudocores for $n=1$, and \Cref{fig:coresb} shows the results for the remaining data sets.
In both cases, we see how the numbers of pseudocores approach one, i.e., the $n$-th order temporal H-index is zero for all nodes, as expected due to~\Cref{theorem:largentozero}.
For the data sets with the three highest numbers of average degree, i.e., \emph{Hospital}, \emph{Malawi}, and \emph{Highschool}, the number of pseudocores decreases slower than for the other data sets. 
The reason is that these data sets allow high numbers of temporal walks.
In the case of the \emph{Malawi} data set, we see a slight increase in the number of pseudocores from $82$ for $n=1$ to $83$ for $n=2$.
This does not conflict with~\Cref{theorem:largentozero} because even though for individual nodes, the $n$-th order temporal H-index approaches zero for increasing $n$, the number of different values of the $n$-th order temporal H-index for the nodes can increase. 

\Cref{alg:streaming} efficiently computes the $i$-th order temporal H-index for all $0\leq i\leq n$ in one pass. 
The result allows choosing a decomposition with a number of cores best suited for a downstream task or application.
Note that we obtain all non-trivial index values for $0\leq n\leq \Delta(\tg)$ and that $\Delta(\tg)$, i.e., the maximal length of a temporal walk, is at most the number of availability times in $\tg$.

\subsection{Temporal Characteristics and Use Case}\label{sec:applications}

We first discuss the temporal characteristics of the $n$-th order temporal H-index. 
Due to the space restrictions, we focus on the outward variant. %
See \Cref{appendix:invsout} for a comparison of node rankings computed with the inward and outward $n$-th order temporal H-indices. 

\Cref{fig:pseudocore} shows an example of the outward $n$-th order temporal H-index applied to the \emph{Malawi} network, which is a human face-to-face contact network~\cite{ozella2021using}.
\Cref{fig:pseudocorea,fig:pseudocoreb,fig:pseudocorec,fig:pseudocored} show the aggregated underlying static graph, and the nodes are colored according to their centrality/core value.
\Cref{fig:pseudocorea,fig:pseudocoreb} show the results for the outward $n$-th order temporal H-index, and $n=32$ and $n=512$, respectively.
The values and range of values are higher for $n=32$ compared to $n=512$, as expected due to \Cref{theorem:largentozero}.
\Cref{fig:pseudocorec,fig:pseudocored} show the values of the static H-index and $k$-core numbers of the aggregated graph $A(\tg)$ for comparison.
First, we note that the $n$-th order temporal H-index differentiates more nodes than the static H-index and $k$-cores.
Respecting the temporal reachability constraints leads to different node rankings compared to the static versions, and the nodes are assigned different (relative) importance. 

More specifically, using the outward $n$-th temporal H-index, nodes with a strong ability to reach other nodes are ranked high. 
To verify this property, we measure the \emph{local and global reachability} of out-pseudocores.
Let $\tg=(V, \tge)$ be a temporal graph, and $r\colon V\times V\rightarrow\{0,1\}$ the indicator function for temporal reachability, i.e.,
$r(u,v)=1$ iff $u$ can reach $v$ via a temporal walk.
We define the global and local \emph{reachability scores} $\rho_g$ and $\rho_\ell$ of a pseudocore $\tg_{(n,k)}=(V',\tge')$ as 

    $\rho_g = \frac{\sum_{u\in V',v\in V}r(u,v)}{|V'|\cdot |V|}$, and $\rho_\ell = \frac{\sum_{u,v\in V'}r(u,v)}{|V'|^2}$, respectively.

\smallskip

\Cref{fig:reachability} shows the global and local reachability scores for the \emph{Malawi} contact network and the \emph{FacebookMsg} communication network.
We show the scores for the $n$-th temporal H-index and the static $k$-core. We choose two values for $n$, where the first equals $2^i$ where $i$ is chosen such that there are at least 20 different ranks, i.e., $n=256$ for \emph{Malawi} and $n=8$ for \emph{FacebookMsg}, and additionally, we choose $2^{i+1}$. 
Furthermore, as an additional baseline, we use the temporal $(k,h)$-core from~\cite{wu2015core} where we choose $h\in\{4,8\}$.
In a $(k,h)$-core, each node has at least $k$ neighbors and at least $h$ temporal edges to each neighbor.
The $x$-axes in \Cref{fig:reachability} show cores by ordered rank from zero to the number of cores minus one, where $x=0$ corresponds to the highest-ranked core.
The results verify that our outward $n$-th order temporal H-index leads to pseudocores characterized by high local and global temporal reachability. 
We discuss the sizes of the cores in \Cref{appendix:sizes}.
The nodes in a high-ranked pseudocore, i.e., with a high $n$-th order H-index, can, on average, reach many of the nodes of the graph (global), and the nodes in the pseudocore (local).
We verify the usefulness of this important property with the following use case.

\subsubsection{Use Case: Influential Spreader Identification}
We evaluate how well the outward $n$-th order temporal H-index captures the node influences of
spreading processes.
Here, we use the stochastic \emph{susceptible-infected-recovered} (SIR) spreading model in temporal networks, which is commonly applied to analyze information or disease dissemination~\cite{pei2015exploring,holme2021fast,zhang2022mitigate}.
We give a detailed description in \Cref{appendix:sir}.
We follow the standard practice of recent works, e.g.,~\cite{lu2016h,liu2018leveraging,li2019identifying,maji2020systematic}, to compare the rankings obtained by centrality measures with the ranking obtained by applying the SIR model. 
We chose two face-to-face (\emph{Malawi} and \emph{Infectious}) and two communication networks (\emph{FacebookMsg} and \emph{Email}) for the experiment.
We chose different infection probabilities $\beta$ and computed for each $\beta$ and node $u\in V$ the mean node influence $R_u$ over 1000 independent SIR simulations leading to the SIR node rankings.
We then compare the SIR rankings with those obtained by the centrality measures using the Kendall $\tau_b$ rank correlation measure~(\Cref{appendix:kendall}).
As baselines, we use the static variants of the \emph{degree centrality}, \emph{H-index}, and \emph{$k$-core}, which are common and popular heuristics for identifying strong spreaders~\cite{lu2016h}.
Furthermore, we used the \emph{temporal closeness centrality}, which was suggested as a strong heuristic for the case of temporal networks in~\cite{oettershagen2020efficient,oettershagen2022computing}, and the recently proposed \emph{temporal walk centrality}~\cite{oettershagen2022temporal}. 
Finally, we use the state-of-the-art heuristics \emph{(local) gravity centrality}~\cite{ma2016identifying,li2021generalized} and extended neighborhood coreness (\emph{coreness+})~\cite{bae2014identifying}, which belong to the best performing $k$-core based heuristics as shown in a recent survey~\cite{maji2020systematic}.
For the $n$-th order temporal H-index (\textsc{Thi}), we set $n=32$ and $n=64$, to include a sufficient neighborhood sizes and to consider sufficient depths of the computation. %
\Cref{fig:sir} shows the results. 
Node rankings obtained from the $n$-th order temporal H-index have,
in almost all cases, a higher mean correlation to the SIR ranking
than the node rankings from the baselines. Only for the
\emph{Malawi} data set, the temporal closeness achieves higher correlations.
In conclusion, the temporal H-index often indicates the
nodes’ capabilities for spreading disease or information better than
the baselines.

\section{Conclusion}
We generalized the static $n$-th order H-index for temporal networks.
We obtained meaningful variants by taking into account the restricted reachability caused by temporal dynamics. 
Our streaming algorithm for the common case of uniform transition times is highly scalable.
In our experiments, we demonstrated the efficiency and effectiveness of our new approaches. 
Finally, we showed that the $n$-th order temporal H-index could successfully identify super-spreaders in a use-case.
In future work, we plan to utilize the pseudocore decomposition to detect cohesive temporal communities and visualize temporal networks.

\section*{Acknowledgements}
This work is funded by the Deutsche Forschungs\-gemein\-schaft (DFG, German Research Foundation) under Germany's Excellence Strategy--EXC-2047/1--390685813.
This research has been funded by the Federal Ministry of Education and Research of Germany and the state of North-Rhine Westphalia as part of the Lamarr-Institute for Machine Learning and Artificial Intelligence, LAMARR22B.
This research is supported by the ERC Advanced Grant REBOUND (834862) and the EC H2020 RIA project SoBigData++ (871042).
Nils Kriege was supported by the Vienna Science and Technology Fund
(WWTF) [10.47379/VRG19009].

\appendix
\section*{Appendix}
We provide the omitted proofs and additional results.
\Cref{table:notation} gives an overview of commonly used notations.
\begin{table*}[ht]
    \caption{Commonly used notations}
    \label{table:notation}
    \centering%
    \resizebox{1\linewidth}{!}{\renewcommand{\arraystretch}{1.0}%
            \begin{tabular}{c@{\hspace{3mm}}l@{}}%
                    \toprule
                    \textbf{Symbol} & \textbf{Definition}
                    \\\midrule
                     $G=(V, E)$ & finite static {graph} with set of nodes $V$ and undirected edges $E\subseteq\{\{u,v\}\subseteq V\mid u\neq v\}$\\
                    $\tg=(V,\tge)$                          & finite temporal graph $\tg$ with nodes $V$ and temp. edges $\tge$ \\
                    $e=(u,v,t,\lambda)$                     & temporal $(u,v)$-edge at time $t$ with transition time $\lambda$\\
                    $\Delta(\tg)$                           & the length of the longest (in \# of edges) temporal walk in $\tg$\\
                    $\maxdeg^-(\tg)$, $\maxdeg^+(\tg)$      & maximal in-degree and out-degrees of $\tg$\\
                    $\hfh{v,-}{n}$                      & outward {$n$-th order temporal H-index} of a node $v\in V$\\
                    $\hfh{v,+}{n}$                      & inward {$n$-th order temporal H-index} of a node $v\in V$\\
${N^-}(u,t)$ & time-dependent in-neighborhood of a node $u$ at time $t$, i.e, $\{(v,u,t_e,\lambda_e)\in \tge \mid t_e+\lambda_e\leq t \}$\\
${N^+}(u,t)$ & time-dependent out-neighborhood of a node $u$ at time $t$, i.e., $\{(u,v,t_e,\lambda_e)\in \tge \mid t_e\geq t \}$\\

$\mathcal{N}^{-}(v,t)$ & all pairs of nodes and starting times $(w,t_w)$ such that there exists a temporal edge $(w,v, t_w, \lambda)\in \tge$ with $t_w+\lambda\leq t$\\

  $\mathcal{N}^{+}(v,t)$ & multiset of pairs of nodes and times $(w,t_w)$ s.t.~there is a temporal edge $(v,w, t', \lambda)\in \tge$ 
with arrival time $t_w=t'+\lambda$ and $t'\geq t$\\
                    $\hop\colon \mathcal{S} \rightarrow \mathbb{N}_0$ & for $S\subseteq\{\!\!\{s\mid s\in\mathbb{N}_0\}\!\!\}$  the max. $i$ s.t.~at least $i$ elements $s\in S$ with $s\geq i$\\
                    $\Gamma^-(v),\Gamma^+(v)$ & inward and outward {reachability trees}\\ 
                     $d(u)$ & depth of $u \in V(\Gamma^\star(v))$ with $\star\in\{-,+\}$\\
                    \bottomrule
            \end{tabular}}
\end{table*}

\section{Omitted Proofs}\label{appendix:proofs}
\begin{proof}[Proof of \Cref{lemma:phihf}]
    We use induction over the distances in the induced subtree $\Gamma^\star_n(v)\subseteq \Gamma(v)$ that only contains nodes $u\in U$ with $d(u)\leq n+1$.
    Our base case considers nodes $\ell\in U$ with $d(\ell)=n$ for which $\phi_n(\ell=(w,t))=|C(\ell)|=\delta^\star(w,t)=\hf{w}{t,\star}{n-d(\ell)}$.
    Now we consider an inner node $u=(w,t)$ of the tree and assume the statement holds for all its children $C(u)$.
    Hence, 
    \begin{align*}
        \phi_n(u)&=\hi{\{\!\!\{ \phi_n(c) \mid c\in C(u)\}\!\!\}}=\hi{\{\!\!\{\hf{w_i}{t_i,\star}{n-d(w_i)} \mid (w_i,t_i)\in C(u) \}\!\!\}}\\&=\hi{\{\!\!\{\hf{w_i}{t_i,\star}{n-(d(u)+1)} \mid (w_i,t_i)\in\mathcal{N} \}\!\!\}}=\hf{w}{t,\star}{n-d(u)}
    \end{align*}
\end{proof}

\begin{proof}[Proof of \Cref{theorem:numdescendants}]
    Due to \Cref{def:hoperator} the root $r$ has at least $k$ children $C(r)$ because $\hf{v}{t,\star}{0}\geq k$.
    For each $(w,t)\in C(r)$ of the children, we can argue that $\hf{w}{t,\star}{1}\geq k$.
    Similar we can argue that in each depth level up to $n$ of $\Gamma^\star(v)$ each node has at least $k$ children.
    Hence, the total number of descendants is at least 
    $
    \sum_{i=1}^{n+1} k^i=\frac{k^{n+2}-k}{k-1}.
    $
\end{proof}

\begin{proof}[Proof of \Cref{theorem:largentozero}]
    Consider the reachability tree $\Gamma^\star(v)$ of the node $v$. 
    Let $d(\Gamma^\star(v))$ be the maximal depths of the tree which is upper bounded by $\Delta(\tg)$.
    Recall that $\Delta(\tg)$ is the temporal diameter of $\tg$, i.e., an upper bound for the maximal number of edges in any walk in $\tg$.
    Then, for $n>d(\Gamma^\star)$ it follows that for each leaf $\ell$ in $\Gamma^\star(v)$, $\phi_n(\ell)=0$ because  $n>d(\Gamma^\star)\geq d(\ell)$. 
    Moreover, the depths of the ancestors $\ell$ in $\Gamma^\star(v)$ is strictly smaller than $d(\ell)$. 
    Hence, the case that $n=d(u)$ in~\Cref{def:treephi} cannot be  reached and due to~\Cref{lemma:phihf} the result follows.
\end{proof}

\begin{proof}[Proof of \Cref{theorem:containment}]
    The first containment property, $\tg_{(n,0)}\supseteq \tg_{(n,1)}\supseteq\ldots\supseteq\tg_{(n,\tau_n-1)}\supseteq\tg_{(n,\tau_n)}$ , follows directly from \Cref{def:nkcore}.
    To prove the second containment property, $\tg_{(0,k)}\supseteq \tg_{(1,k)}\supseteq\ldots\supseteq\tg_{(\eta_k-1,k)}\supseteq\tg_{(\eta_k,k)}$, let $\Gamma^\star(v)=(U,E,r)$ be a reachability tree of any node $v\in V$.
    We show that $\phi_n(r)\geq \phi_{n+1}(r)$ which together with \Cref{lemma:phihf} proves the statement. 
    Let $p\in U$ be a node $\Gamma^\star(v)$ with $d(p)=n$ and therefore $\phi_n(p)=|C(p)|$  which is an upper bound on $\phi_n(p)$ for all $n\in \mathbb{N}$.
    Moreover, we have $\phi_{n+1}(p)=\hi{\{\!\!\{|C(p)| \mid c\in C(p)\}\!\!\}}$ with $C(p)$ being the set of children of $p$.
    And, $\hi{\{\!\!\{|C(p)| \mid c\in C(p)\}\!\!\}}$ is also upper bounded by $|C(p)|$. Hence, $\phi_{n}(p)\geq\phi_{n+1}(p)$.
    We now assume the statement holds for all children $c\in C(u)$ of some node $u\in U$.
    By induction, it follows that 
    $\phi_{n}(u)=\hi{\{\!\!\{\phi_{n}(c) \mid c\in C(u)\}\!\!\}}\geq \hi{\{\!\!\{\phi_{n+1}(c) \mid c\in C(u)\}\!\!\}}=\phi_{n+1}(u)$.
    Hence, the result follows.
\end{proof}

Before we prove \Cref{theorem:algstream}, we introduce the following lemmas.
\begin{lemma}\label{lemma:streamdeg}
    Let $t_\ell$ be the latest time stamp such that \Cref{alg:streaming} processed all edges with availability time $t_\ell$. 
    At any iteration of the for-loop in line~\ref{alg:streaming:forloop}, when processing the temporal edge $e=(u,v,t)$,
    it holds $deg[u]=\delta^{+}(u,t_\ell)$ and $deg[v]=\delta^{+}(v,t_\ell)$.
\end{lemma}
\begin{proof}%
    \Cref{alg:streaming} uses $lt[u]$ to store the last time stamp of an edge leaving node $u$.
    If the time stamp $t$ of the currently processed edge $e=(u,v,t)$ is earlier than $lt[u]$, i.e., $lt[u]>t$, 
    the algorithm updates the values $lt[u]$ and $deg[u]$ by setting $lt[u]=t$ and $deg[u]=\pi[u][0].length$.
    This only happens if all edges with availability times later than $t$ are processed due to the processing of 
    the edges in reverse chronological order.
    At this point and due to line~\ref{alg:streaming:updatepione}, the length of $\pi[u][0]$ equals 
    the number of edges leaving $u$ up to time $t$ (not including the current edge $e$).
    Hence, $deg[u]=\delta^{+}(u,t_\ell)$ until the next update. %
    Similarly, the result holds for $deg[v]$. 
\end{proof}

\begin{lemma}\label{lemma:algstreamlem}
    After each iteration of the for loop in line~\ref{alg:streaming:forloop}, processing edge $e=(u,v,t)$,
    the lists $\pi[w][j]$ contain the values $X=\{\!\!\{(t',\hf{y}{t'}{j})\}\!\!\}$ such that $\hop_{t}(X)=\hf{w}{t}{j}$ for all $1\leq j \leq n$ and all $w\in V$.
\end{lemma}
\begin{proof}%
    Let $e_1,\ldots,e_m$ be the sequence of temporal edges in reverse chronological order (ties broken arbitrarily).
    After processing the first edge $e_1=(u_1,v_1,t_1)$ the statement holds with the lists $\pi[w][j]$ empty and $\hf{w}{t_1}{j}=0$ for $1\leq j \leq n$ and all $w\in V$.
    Now assume the statement holds after the $(k-1)$-th edge, and we consider the $k$-th edge.
    When $e_k=(u_k,v_k,t_k)$ arrives all edges with time stamps larger than $t_k$ have been processed.
    For $j=1$, by our assumption $\pi[u_k][j]$ contains pairs $(t_i, d_i)$ for $t_i\geq t_k$ such that $v_i$ is a neighbor that has a temporal degree $\delta^+(v_i,t')=d_i$ for some $t'\geq t_i$. Now after adding $(t_k,deg[v_k])$, it follows that
    $\hop_{t_k}(\pi[u_k][1])=\hop(\{\!\!\{d_i\mid (t', d_i)\in\pi[u_k][j]\wedge t'+\lambda\geq t_k\}\!\!\})=\hf{u_k}{t_k}{1}$.
    Similarly, for $1\leq j \leq n$, in line~\ref{alg:streaming:forloopn} and following, the algorithm appends to each list $\pi[u_k][j+1]$ the pair $p=(t_k, \hop_{t_k}(\pi[v_k][j]))$. 
    By applying our assumption, we can rewrite $p=(t_k, \hf{v_k}{t_{k}}{j})$. 
    And, $\hop_{t_k}(\pi[u_k][j+1])=\hop(\{\!\!\{\hf{v_k}{t_{k}}{j}\mid (t', \hf{v_k}{t_{k}}{j})\in\pi[u_k][j]\wedge  t'+\lambda\geq t_k\}\!\!\})=\hf{u_k}{t_k}{j+1}$.
\end{proof}
\begin{proof}[Proof of \Cref{theorem:algstream}]
    The algorithm terminates after iterating over the edges and nodes. The correctness follows from~\Cref{lemma:streamdeg,lemma:algstreamlem}.
    Note for the case of $n=0$, we only need to set $h^{(0)}_v$ to the length of $\pi[v][0]$ (see line~\ref{alg:streaming:nzero}) because due to line~\ref{alg:streaming:updatepione} the list contains one entry for each outgoing edge from $v$.
    
    For the running time, initialization is done in $\mathcal{O}(|V|\cdot n)$ and we iterate over the edges once.
    In each iteration of the for loop in line \ref{alg:streaming:forloop}, we have another loop with $n$ iterations (line \ref{alg:streaming:forloopn}). During this second loop, we compute $\hit{t}{\pi[v][j]}$. The running time of $\hit{t}{\pi[v][j]}$ is linear in the length of the list $\pi[v][j]$ which is upper bounded by the out-degree of node $v$. 
    Therefore, for the first for loop a running time of $\mathcal{O}(|\tge|\cdot n \cdot \maxdeg)$ follows.
    The running time of lines \ref{alg:streaming:forloopfinalstart}--\ref{alg:streaming:forloopfinalend} 
    is similarly in $\mathcal{O}(|V|\cdot n \cdot \maxdeg)$.
\end{proof}
\begin{figure}\centering
    \begin{subfigure}{0.5\linewidth}\hfill%
        \includegraphics[width=\linewidth]{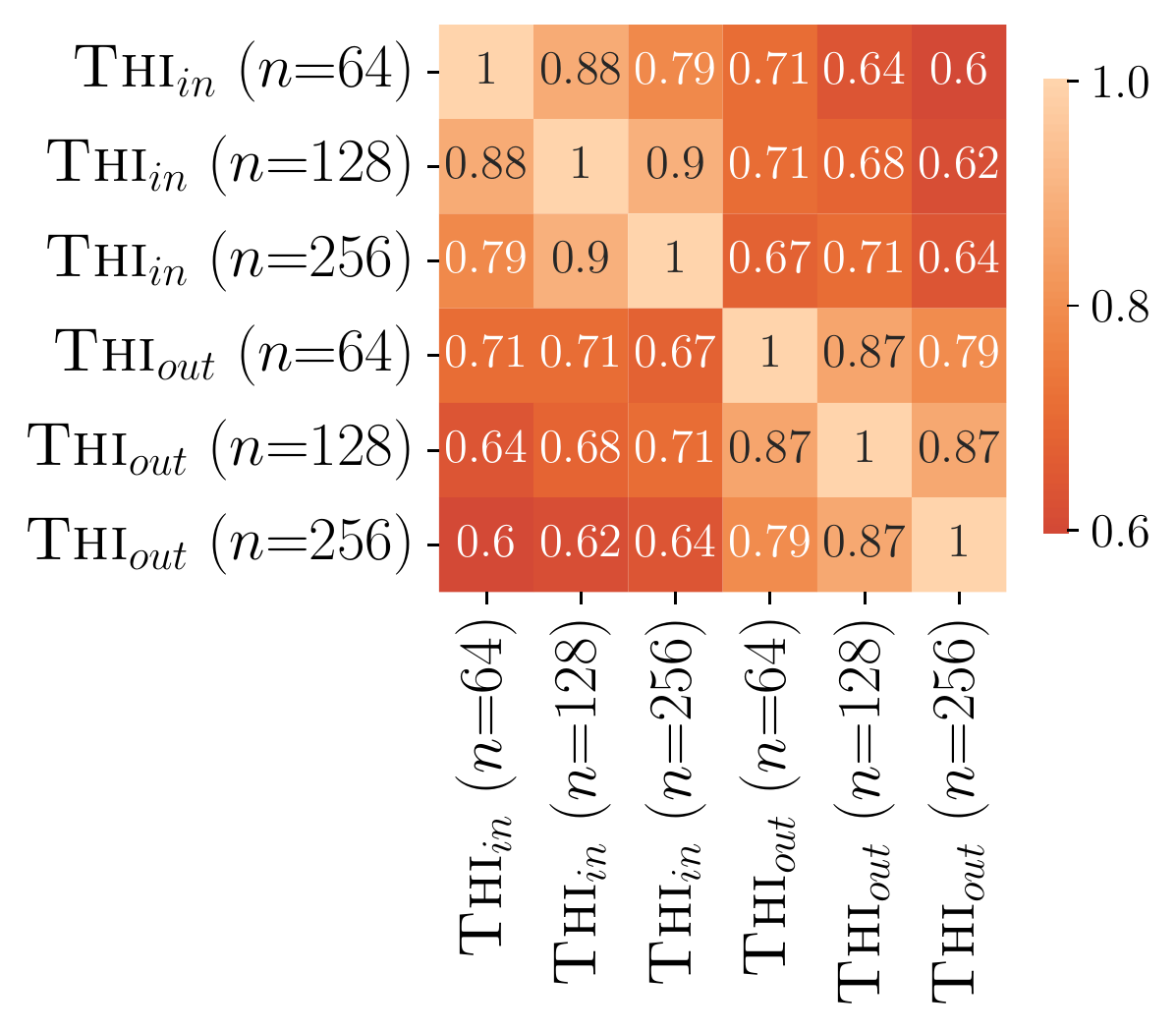}
        \caption{\emph{Malawi}}
        \label{fig:invsouta}
        \vspace{-3mm}
    \end{subfigure}\hfil%
    \begin{subfigure}{0.5\linewidth}\hfill%
        \includegraphics[width=\linewidth]{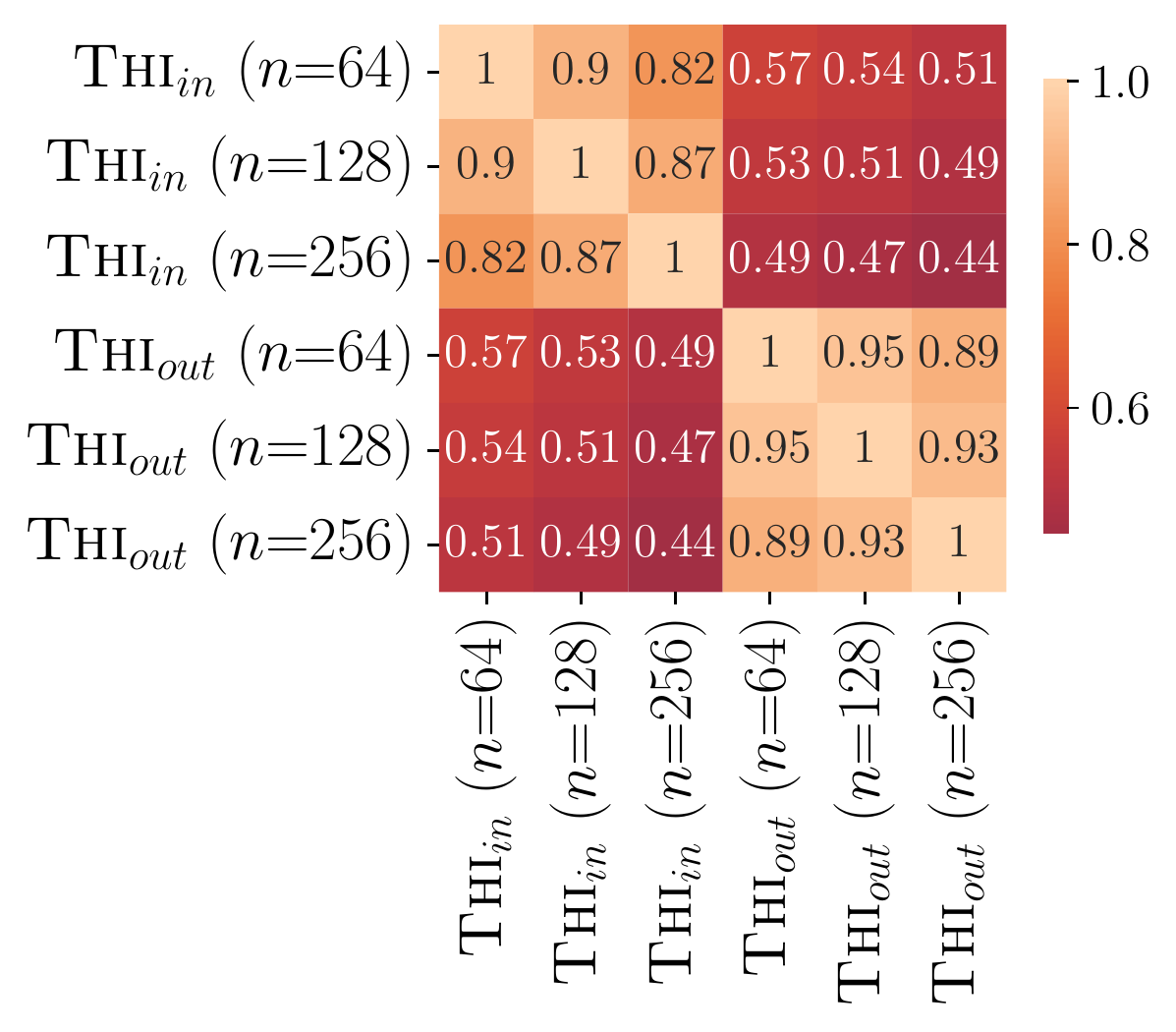}
        \caption{\emph{FacebookMsg}}
        \label{fig:invsoutb}
        \vspace{-3mm}
    \end{subfigure}
    \caption{Kendall $\tau_b$ correlation matrices of the rankings using the inward and outward $n$-th order temporal H-index.}
    \label{fig:invsout}
    \vspace{-3mm}
\end{figure}

\section{Computation}\label{appendix:computation}

In this section, we describe the naive recursive algorithm and the computation of the inward $n$-th order temporal H-index by reversing the temporal graph.

\subsection{Naive Recursive Algorithm}\label{appendix:naive}

The recursive definition of the temporal $n$-th order H-index enables a straightforward top-down approach.
We provide the pseudocode in \Cref{appendix:computation}. 
\Cref{alg:topdown} shows a recursive algorithm for computing $\hfh{v,\star}{n}$ for $\star\in\{-,+\}$ and all $v\in V$, utilizing memoization to avoid recomputation of already computed values of $\hf{v}{t,\star}{n}$.
\begin{theorem}\label{theorem:algrec}
    Let $\star\in \{-,+\}$, \Cref{alg:topdown} computes $\hfh{v,\star}{n}$ for all $v\in V$ 
    in time $\mathcal{O}(|V|\cdot n\cdot (\maxdeg^\star)^2)$ with space $\mathcal{O}(|V|\cdot n\cdot \maxdeg^\star)$.
\end{theorem}
\begin{proof}[Proof of \Cref{theorem:algrec}]
    The running time and space complexity depend on the total size of $m[v][i]$ which consists of $|V|\cdot n$ associative arrays, which can be implemented using perfect hashing as the time steps of the incoming edges at each node are known in advance.
    Each of the associative arrays contains at most $\maxdeg$ entries, where $\maxdeg$ is the maximum degree in $\tg$, and the space complexity follows.
    To obtain each entry of the hash maps, we have to compute $\hop(L)$
    in $\mathcal{O}(\maxdeg)$ time because the size of $L$ is upper bounded by $\maxdeg$, leading to the stated running time complexity.
    
    The algorithm terminates due to line~\ref{alg:topdown:nzero} and because in each call of the function $R$ the value of parameter $n$ is decreased (line~\ref{alg:topdown:ndecrease}).
    We show that $R(v,n,t)=\phi_n(u=(v,t))$ in the reachability tree $\Gamma_n(v)=(U,E)$ that contains all nodes $u\in U$ with depth $d(u)\leq n+1$
    We use induction over the recursion depths.
    The maximum depth of the recursion tree is $n$ due to lines~\ref{alg:topdown:nzero} and~\ref{alg:topdown:ndecrease}.
    The base case is for depths $n$ where we have $R(v,0,t)=\delta(v,t)=\phi_n(u=(v,t))$.
    Now, for the case of depths $j<n$, we consider $R(v',n-j,t')$ and assume the statement holds for depths larger than $j$.
    Due to lines~\ref{alg:topdown:innerfor}f, and by applying the induction hypothesis it follows that
    $L=\{\!\!\{R({w, n-1, t+\lambda})\mid (v,w,t,\lambda)\in\tge \}\!\!\}=\{\!\!\{\phi_n(c)|c\in C(u)\}\!\!\}$, where $C(u)\subset U$ is the set of children of the node $u=(v',t')\in U$ in $\Gamma_n(v)$. 
    Therefore, $R(v',n-j,t')=\hop(L)=\phi_n(u=(v',t'))$.
    With~\Cref{lemma:phihf} the result follows.
\end{proof}
\begin{algorithm}[h]\small
    \label[algorithm]{alg:topdown}
    \caption{Algorithm \textsc{Recurs} using memoization}
    \Input{Temporal graph $\tg=(V,\tge)$, $n\in\mathbb{N}$}
    \Output{$\hfh{v,\star}{n}$ for all $v\in V$}
    Initialize empty array of associative arrays $m[v][i]$ for all $v\in V$ and $i\in[n]$\;
    \BlankLine
    \SetKwProg{Fn}{Function}{:}{}
    \SetKwFunction{rec}{R}
    \Fn{\rec{$v$, $n$, $t$}}{
        \If{$m[v][n]$ has entry for $t$}{\label{alg:topdown:memoization}
            \Return $m[v][n][t]$
        }
        \If{$n = 0$ \KwOr $\tdegstar{v}{t}=0$}{\label{alg:topdown:nzero}
            \Return $\tdegstar{v}{t}$
        }
        Initialize empty multiset $L$\;
        \ForEach{$e=(v,w,t,\lambda)$ with $(w,t)\in \mathcal{N}^\star(v,t)$}{\label{alg:topdown:innerfor}
            $L \gets L \cup \{ $\rec{$w$, $n-1$, $t+\lambda$}$ \}$\;\label{alg:topdown:ndecrease}
            (or $L \gets L \cup \{ $\rec{$w$, $n-1$, $t$}$ \}$ if $\star=-$) \;
        }
        
        $m[v][n][t]\gets \hi{L}$\;
        \Return $m[v][n][t]$\;
        
    }
    \BlankLine  
    
    Initialize $h^{(n)}_{v,\star}=0$ for all $v\in V$\;
    \ForEach{$v\in V$}{
        $h^{(n)}_{v,\star}=\rec{v, n, 0}$\;
    }
    \Return $h^{(n)}_{v,\star}$ for all $v\in V$\;
\end{algorithm}
\subsection{Reverse Computation}\label{appendix:revcomp}
In order to compute the inward $n$-th order temporal H-index using \Cref{alg:streaming}, we first transform the temporal network $\tg$ using
the \emph{temporal transpose} $\mathcal{T}(\tg)$ introduced in~\cite{oettershagen2020efficient}.
\begin{definition}[Temporal transpose~\cite{oettershagen2020efficient}]
    For a temporal graph $\tg=(V,\tge)$, we call $\mathcal{T}(\tg)=(V,\tilde{\tge})$ with edge set $\tilde{\tge} = \{(v,u,t_{max}-t,\lambda)\mid(u,v,t,\lambda)\in \tge\}$ and $t_{max}=\max\{t+\lambda\mid(u,v,t,\lambda)\in \tge\}$ the \emph{temporal transpose} of $\tg$.
\end{definition} 
Computing the temporal transpose is possible in $\mathcal{O}(m)$ time and space.
We use the following helpful result.
\begin{proposition}\label{lemma:reverse}
    Let $\tg$ be a temporal graph with equal transition times for all $e\in\tge$ and $\mathcal{T}(\tg)$ it temporal transpose.
    Then, $\hfh{v,+}{i}$ in $\mathcal{T}(\tg)$ equals $\hfh{v,-}{i}$ in $\tg$ for all $0\leq i \leq n$ and $v\in V$.
\end{proposition}
\begin{proof}
    The proposition follows from Lemma 4 in \cite{oettershagen2022computing}.
\end{proof}
Due to \Cref{lemma:reverse}, we can efficiently compute $\hfh{v,-}{i}$ for all $0\leq i \leq n$ and $v\in V$ by first computing the temporal transpose $\mathcal{T}(\tg)$ and then call \Cref{alg:streaming} with $\mathcal{T}(\tg)$.

\begin{table}
   \centering
   \vspace{1mm}
   \caption{Overview of related centrality measures and core decompositions, with $\mathcal{T}=\{t\mid (u,v,t,\lambda)\in \tge\}$, $\tau_{max}$ the largest cardinality of availability or arrival times at a node, $T$ the total length of the time interval spanned by the temporal graph, and $I$ the length of a time window. The time complexities are for computing the centrality or core number all nodes.}
   \label{tab:overview2}
   \resizebox{\linewidth}{!}{\renewcommand{\arraystretch}{1.2}\setlength{\tabcolsep}{13pt}
           \begin{tabular}{lccc}\toprule
                   \textbf{Method}   & \textbf{Running Time} & \textbf{Space}  & \textbf{Reference} \\ \midrule
                   Static H-index       & $\mathcal{O}(|V|\cdot \maxdeg)$ & $\mathcal{O}(|V|)$ & \cite{hirsch2005index}\\
                   Temporal Closeness   & $\mathcal{O}(|V|^2 + |V||\tge|)$ & $\mathcal{O}(|V| + |\tge|)$ & \cite{wu2014path}\\
                   Temporal Betweenness & $\mathcal{O}(|V|^3\cdot \mathcal{T}^2)$ & $\mathcal{O}(|V|\cdot \mathcal{T} + |\tge|)$ & \cite{BussMNR20}\\
                   Temporal PageRank/Katz   & $\mathcal{O}(|\tge|)$ & $\mathcal{O}(|V|)$ & \cite{RozenshteinG16,katztg}\\
                   Temporal Walk Centrality & $\mathcal{O}(|\tge|\cdot \tau_{max})$ & $\mathcal{O}(|V|\cdot \tau_{max})$ & \cite{oettershagen2022temporal}\\ \midrule
                   Static $k$-Core       & $\mathcal{O}(|V|+|E|)$ & $\mathcal{O}(|V|)$ & \cite{lu2016h} \\
                   Temporal $(k,h)$-Core & $\mathcal{O}(|V|\cdot \maxdeg)$ & $\mathcal{O}(|V|)$ & \cite{wu2015core}\\
                   Temporal Span-Core    & $\mathcal{O}(\tge\cdot T^2)$ & $\mathcal{O}(|\tge|+T)$ & \cite{galimberti2020span}\\
                   \bottomrule
               \end{tabular}
       }	
\end{table}
\begin{figure*}\centering%
    \begin{subfigure}{0.248\linewidth}\hfill%
        \includegraphics[width=\linewidth]{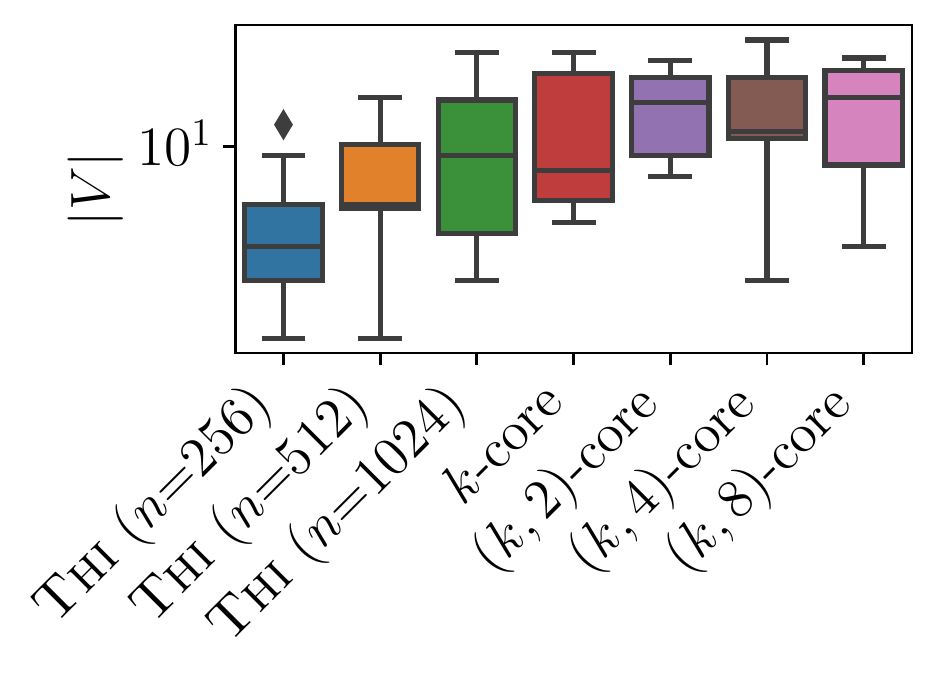}
        \caption{\emph{Malawi}}
        \label{fig:sizesa}
    \end{subfigure}%
    \begin{subfigure}{0.248\linewidth}\hfill%
        \includegraphics[width=\linewidth]{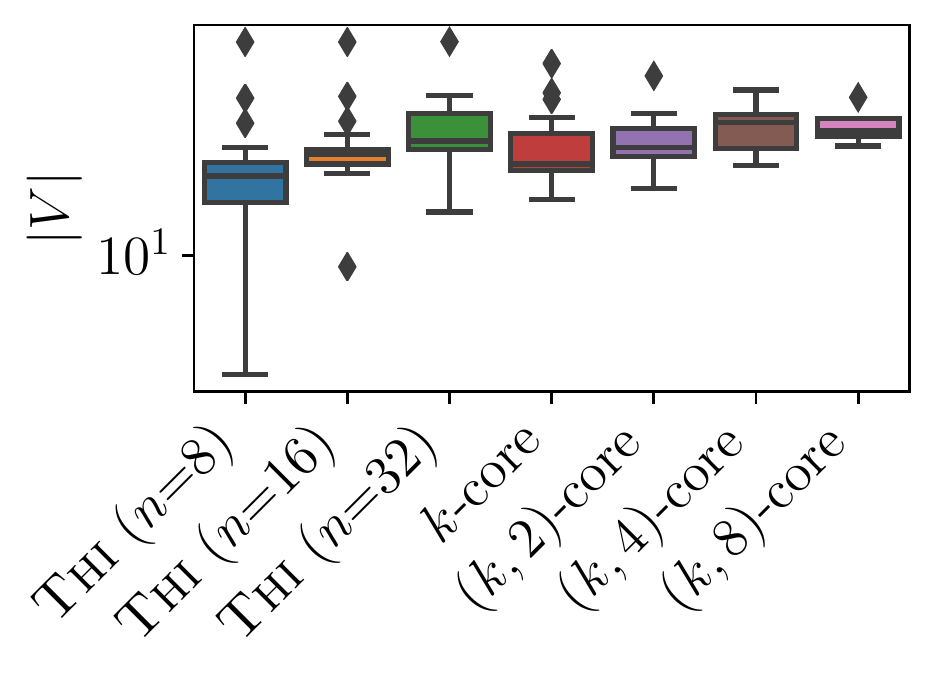}
        \caption{\emph{FacebookMsg}}
        \label{fig:sizesb}
    \end{subfigure}
    \begin{subfigure}{0.248\linewidth}\hfill%
        \includegraphics[width=\linewidth]{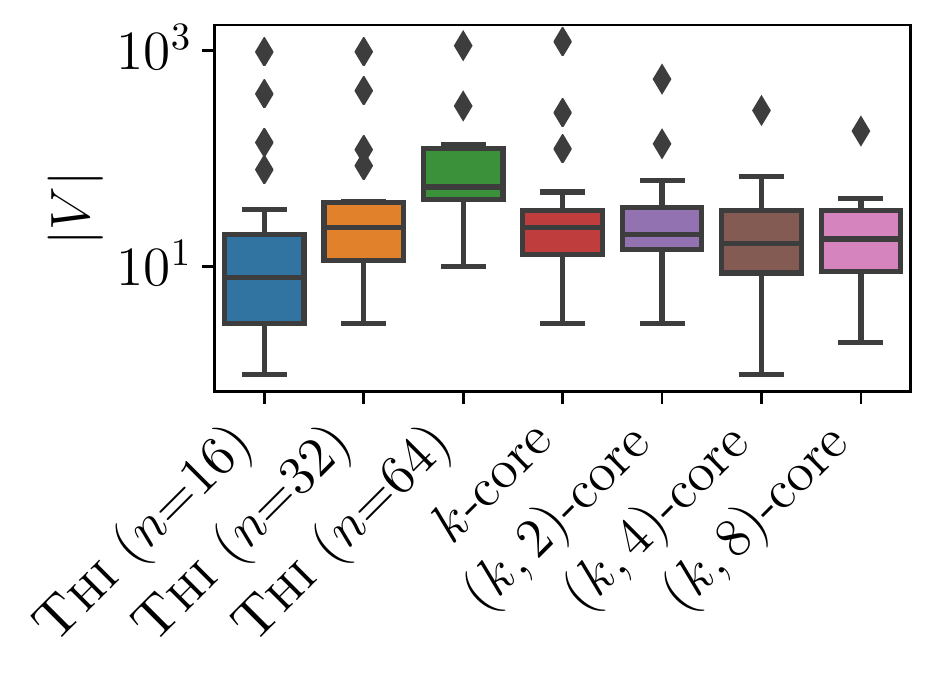}
        \caption{\emph{Email}}
        \label{fig:sizesc}
    \end{subfigure}%
    \begin{subfigure}{0.248\linewidth}\hfill%
        \includegraphics[width=\linewidth]{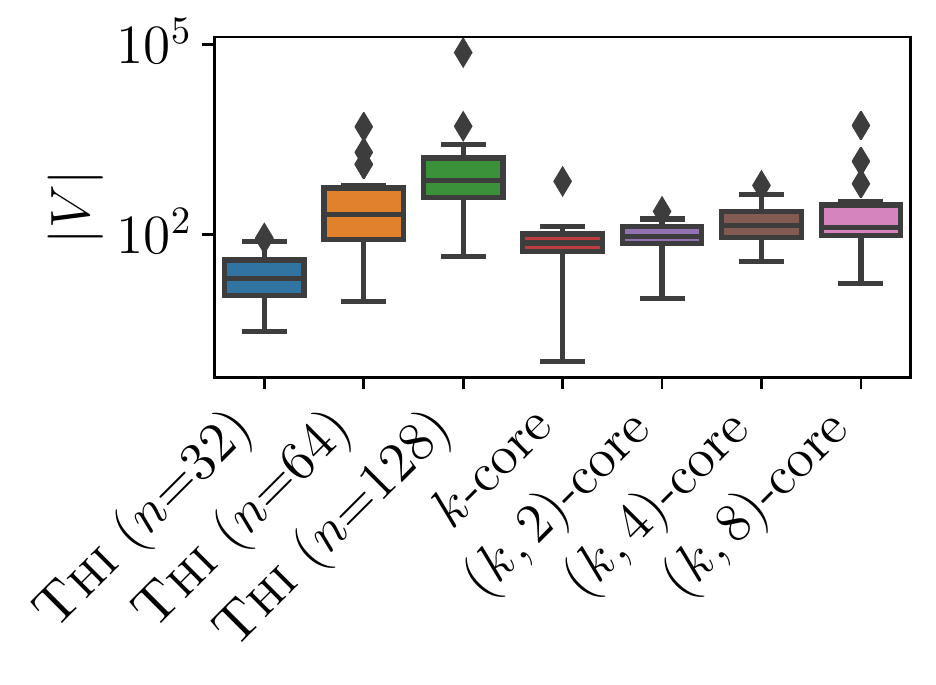}
        \caption{\emph{Enron}}
        \label{fig:sizesd}
    \end{subfigure}
    \vspace{-4mm}
    \caption{Box-plots of the number of vertices in the computed cores. The $y$-axes are logarithmic.}
    \label{fig:sizes}
\end{figure*}

\section{Experimental Evaluation}
We provide additional experimental results and the details.
\subsection{Kendall Rank Correlation}\label{appendix:kendall}
The Kendall rank correlation coefficient is a common and standard way of comparing node rankings and centrality measures~\cite{grando2016analysis}.
Because node rankings may contain ties, we use the Kendall $\tau_b$ rank correlation variant that makes adjustments for ties. For the formal definition, please refer to~\cite{kendall1938new}.
The Kendall $\tau_b$ rank correlation takes on values between one and minus one: values close to one indicate similar rankings, close to zero no correlation, and close to minus one a strong negative correlation.

\subsection{Inward vs Outward $n$-th Temporal H-Index}\label{appendix:invsout}
\Cref{fig:invsout} shows the Kendall $\tau_b$ correlations between the node rankings obtained by the inward and outward $n$-th order temporal H-index for $n\in\{64,128,256\}$.
The correlation between the rankings using inward or outward $n$-th order temporal H-index for different $n$ is high, which we expected.
The correlation between rankings obtained from the inward and the outward $n$-th order temporal H-index is lower.
However, we still observe correlations of at least $0.6$ (\emph{Malawi}) and $0.44$ (\emph{FacebookMsg}) as there 
are nodes with high (or low) inward and outward $n$-th order temporal H-index in the data sets.

\subsection{Sizes of the Pseudocores}\label{appendix:sizes}
\Cref{fig:sizes} shows boxplots of the vertex set sizes $|V|$ of the cores for the temporal H-index, the static $k$ core, and the temporal $(k,h)$-cores. We show the results for \emph{Malawi}, \emph{FacebookMsg}, \emph{Email}, and \emph{Enron}.
For the temporal H-index, the pseudocore sizes increase with increasing $n$ because the number of pseudocores decreases (\Cref{theorem:largentozero}). Hence, the sizes are controllable using the parameter~$n$.

\subsection{SIR Model}\label{appendix:sir}
The stochastic \emph{susceptible-infected-recovered} (SIR) spreading model is a standard model for analyzing spreading processes~\cite{pei2015exploring,holme2021fast}.
In the model, each node is either \emph{susceptible}, \emph{infected}, or \emph{recovered}. 
We used the SIR model as described in~\cite{holme2021fast} and the implementation for temporal networks provided by the author.
In the beginning, a single node $u\in V$ is infected.
The infection can spread along temporal edges with infection probability $\beta$.  
An infected node will recover after time $\delta$ sampled with probability $P(\delta)=\nu \exp(-\nu\delta)$ from an exponential distribution, where $\nu$ is set to 20 time units of the network.
At the start of the simulation, a single node is infected.
The node influence $R_u$ of the initially infected node $u\in V$ is the number of nodes that are 
either infected or recovered after the simulation finishes, i.e., when no further infections are possible.

\end{document}